\documentclass[a4paper,fleqn]{aa}
\pdfoutput=1
\usepackage[english]{babel}
\usepackage{inputenc}
\usepackage{amsmath, amssymb}
\usepackage{graphicx}
\usepackage{multicol}
\usepackage{multirow}
\usepackage{booktabs, dcolumn}
\usepackage[draft]{hyperref}

\usepackage[varg]{txfonts}

\newcolumntype{d}[1]{D{.}{.}{#1}}
\newcounter{refcount}
\newcommand{\prefcount}{\stepcounter{refcount}\therefcount}
\newcommand{\pprefcount}{(\stepcounter{refcount}\therefcount)~}
\newcommand{\deriv}[2]{\ensuremath{\frac{\partial {#1}}{\partial {#2}}}}
\newcommand{\median}[1]{\ensuremath{\mathcal{M}\!\left( #1 \right)}}
\newcommand{\fracdelta}[1]{\ensuremath{\frac{#1}{\sigma_{m_\lambda}^2}}}
\newcommand{\feh}{[\rm{Fe/H}]}
\newcommand{\met}{[\rm{M/H}]}
\newcommand{\teff}{\ensuremath{T_{\rm{eff}}}}
\newcommand{\mass}{\ensuremath{M}}
\newcommand{\bX}{\ensuremath{\textbf{X}}}
\newcommand{\bO}{\ensuremath{\textbf{O}}}
\newcommand{\dtau}{\delta_{\tau,j}}
\newcommand{\dfeh}{\delta_{\feh,j}}
\newcommand{\dM}{\delta_{M,j}}
\newcommand{\MAD}{\ensuremath{\rm{M.A.D.}}}
\newcommand{\Msun}{\ensuremath{\rm{M}_\odot}}
\newcommand{\threeparams}{$\teff, \log g$, and $\feh$}
\newcommand{\stud}{Student's t-distribution}
\newcommand\MyHead[2]{%
  \multicolumn{1}{c}{\parbox{#1}{\centering #2}}
}
\newcommand{\myimage}[1]{\begin{center}\includegraphics[angle=0, width=0.9\textwidth]{#1}\end{center}}
\newcommand{\myimageH}[1]{\begin{center}\includegraphics[angle=0, height=0.95\textheight]{#1}\end{center}}

\newcommand{\myimagesmall}[1]{\begin{center}\includegraphics[angle=0, width=0.48\textwidth]{#1}\end{center}}
\newcommand{\myimages}[2]{\begin{center}\includegraphics[angle=0, width=#2\textwidth]{#1}\end{center}}

\title{A Unified tool to estimate Distances, Ages, and Masses (UniDAM) from spectrophotometric data.\thanks{The unified tool source code is available at \url{https://github.com/minzastro/unidam}, tables with results are available in electronic form
at the CDS via anonymous ftp to cdsarc.u-strasbg.fr (130.79.128.5)
or via \url{http://cdsweb.u-strasbg.fr/cgi-bin/qcat?J/A+A/}}}
\titlerunning{Stellar parameters}
\author{Alexey Mints\inst{\ref{inst1}, \ref{inst2}}\thanks{email: mints@mps.mpg.de} \and Saskia Hekker\inst{\ref{inst1}, \ref{inst2}}}
\authorrunning{A. Mints and S. Hekker}
\institute{Max Planck Institute for Solar System Research, Justus-von-Liebig-Weg 3, 37077 Göttingen, Germany \label{inst1} \and
Stellar Astrophysics Centre, Department of Physics and Astronomy, Aarhus University, Ny Munkegade 120, DK-8000 Aarhus C, Denmark \label{inst2}}

\abstract {Galactic archaeology, the study of the formation and evolution of the Milky Way by reconstructing its past from its current constituents, requires precise and accurate knowledge of stellar parameters for as many stars as possible. To achieve this, a number of large spectroscopic surveys have been undertaken and are still ongoing. } 
{So far consortia carrying out the different spectroscopic surveys have used different tools to determine stellar parameters of stars from their derived effective temperatures ($T_{\rm eff}$), surface gravities ($\log g$), and metallicities ([Fe/H]); the parameters can be combined with photometric, astrometric, interferometric, or asteroseismic information. Here we aim to homogenise the stellar characterisation by applying a unified tool to a large set of publicly available spectrophotometric data.} 
{We used spectroscopic data from a variety of large surveys combined with infrared photometry from 2MASS and AllWISE and compared these in a Bayesian manner with PARSEC isochrones to derive probability density functions (PDFs) for stellar masses, ages, and distances. We treated PDFs of pre-helium-core burning, helium-core burning, and post helium-core burning solutions as well as different peaks in multimodal PDFs (i.e. each unimodal sub-PDF) of the different evolutionary phases separately.} 
{For over 2.5 million stars we report mass, age, and distance estimates for each evolutionary phase and unimodal sub-PDF. We report Gaussian, skewed, Gaussian, truncated Gaussian, modified truncated exponential distribution or truncated Student's t-distribution functions to represent each sub-PDF, allowing us to reconstruct detailed PDFs. Comparisons with stellar parameter estimates from the literature show good agreement within uncertainties.} 
{We present UniDAM, the unified tool applicable to spectrophotometric data of different surveys, to obtain a homogenised set of stellar parameters.}


\date{XXX/YYY}

\begin{document}

\let\subsectionautorefname\sectionautorefname
\let\subsubsectionautorefname\sectionautorefname
\maketitle\ . 

\section{Introduction}\label{sec:intro}
The Milky Way Galaxy is a unique object to test our understanding of stellar evolution, galaxy formation, and cosmology. For this test a detailed map of our Galaxy, including bulge, disk, halo, spiral structure, and streams formed by recent mergers, is required. Through an analysis of the Galaxy, we can learn how our Galaxy has formed, evolved, and how it interacts with its surroundings.
To build such a map we need to find the distribution of stars in their positions, velocities, chemical compositions, and ages throughout the Galaxy.
These parameters can be measured with different kinds of observations, such as astrometry, photometry, spectroscopy, and asteroseismology.

Astrometric observations provide stellar positions, proper motions and, through parallaxes, distances. These kind of data have been available for decades \citep[see e.g.][]{1970ROAn....5.....W, 1991adc..rept.....G, 1997A&A...323L..49P}. In the nearest future Gaia \citep{2016arXiv160904172G} will vastly increase the precision and amount of such information. The first Gaia data release \citep{2016arXiv160904303L} already provides proper motions and parallaxes for about two million stars, although the precision of parallaxes in this sample limits their application \citep[see e.g.][]{2016arXiv160905390S}. In the next data releases Gaia will provide high-precision parallaxes and proper motions for hundreds of millions of stars, vastly increasing our knowledge of the Galaxy. 

Another rich source of data is spectroscopy, which can provide radial velocities, chemical compositions as well as the effective temperature $\teff$ and the surface gravity $\log g$.
These data can be used to derive stellar ages and distances (see below). A growing number of large spectroscopic surveys, such as RAdial Velocity Experiment \citep[RAVE;][]{2016arXiv160903210K}, Large Sky Area Multi-Object Fibre Spectroscopic Telescope (LAMOST) surveys \citep{2015RAA....15.1095L}, Apache Point Observatory Galactic Evolution Experiment \citep[APOGEE;][]{2014ApJS..211...17A}, Sloan Extension for Galactic Understanding and Exploration \citep[SEGUE;][]{2009AJ....137.4377Y}, and Gaia-ESO \citep{GAIA_ESO} provide rich spectroscopic information for millions of stars. 

Photometric surveys can be used in two ways. Stromgren or Washington-DDO51 photometry can be used to estimate stellar parameters such as the effective temperature $\teff$, surface gravity $\log g$, and metallicity $\feh$ \citep[see][]{2011AA...530A.138C}, or to separate giants from dwarfs for spectroscopic follow-up \citep{2000AJ....120.2550M}. Otherwise, broadband photometry is commonly used as a supplement to spectroscopic data to infer stellar distances.

Asteroseismology is a relatively young and very promising method of exploring stars. 
For low-mass dwarfs, subgiants, and red giant stars asteroseismology can provide a direct measure of mean density and surface gravity.
The surface gravities measured by asteroseismic methods have much higher precision than spectroscopic methods.
In case the effective temperature $\teff$ or luminosity $L$ are also measured it is possible to obtain stellar mass and radius from the asteroseismic observables. When compared with models stellar ages can also be determined from asteroseismology.
COnvection ROtation and planetary Transits \citep[CoRoT;][]{2006ESASP1306...33B}, \textit{Kepler} \citep{2010Sci...327..977B}, and \textit{K2} \citep{2014PASP..126..398H} observations provide such asteroseismic data. These datasets provide high-precision data on small patches of the sky and also have proved to be a perfect sample for the calibration of large spectroscopic surveys \citep[see e.g.][and Tayar et al. 2017, accepted]{2016A&A...594A..43H, 2016AJ....152....6W}. Transiting Exoplanet Survey Satellite \citep[TESS;][]{2014SPIE.9143E..20R} and PLAnetary Transits and Oscillations of stars \citep[PLATO;][]{2014ExA....38..249R} space missions are scheduled to be launched in 2018 and 2024, respectively, and will vastly increase the number of stars with asteroseismic data in the coming years. 

Stellar ages and distances remain among the most challenging parameters to measure. A comprehensive list of age determination methods is given in \cite{2010ARA&A..48..581S}. For a number of stars ages can be derived from asteroseismic observations \citep{2016AN....337..823S} or from carbon and nitrogen abundances \citep{2016MNRAS.456.3655M}. When these data are not available, a typical approach is to compare the parameters directly derived from spectroscopic measurements, which we designate as observed parameters, such as the effective temperature $\teff$, surface gravity $\log g$, and  metallicity $\feh$, to a grid of stellar models. A model or a set of models that have their parameters close to observed parameters give estimates of ages, masses, and absolute magnitudes $M_\lambda$ of a star. Then by comparing the absolute magnitudes to visible magnitudes $m_\lambda$ from photometric surveys we can estimate distances to stars. An overview of this approach is given by \cite{2005A&A...436..127J} and \cite{2011A&A...532A.113B}.

Proper application of this approach requires some care. First, the transformation from observed (\threeparams) to stellar parameters (age and mass) and distance is often degenerate, with the same observables giving two or more possible combinations of stellar parameters. This degeneracy can in some rare cases be resolved when additional observables are available, for example from asteroseismology. Second, the interstellar extinction needs to be accounted for in distance estimations. Extinction values can be taken from external sources or can be derived from observables. Both ways have their advantages and disadvantages. We discuss this in Section \ref{sec:bayes}. Third, observed parameters have their uncertainties and correlations that have to be propagated to uncertainties in stellar parameters.

In the literature a number of methods based on the comparison of observed parameters from spectroscopic and photometric surveys with models to estimate distances and other stellar parameters were proposed and used recently. Here we briefly discuss some of them, and how they deal with the issues stated above. 

\paragraph*{GCS.}
The Geneva-Copenhagen Survey (GCS) \citep{2011AA...530A.138C} team exploited the advantage of having \textit{HIPPARCOS} parallaxes for the majority of their objects; this facilitated the calculation of absolute magnitudes for each star. \citet{2011AA...530A.138C} used a Bag of Stellar Tracks and Isochrones (BASTI) \citep[][and references therein]{2009ApJ...697..275P} and PAdova and TRieste Stellar Evolution Code (PARSEC) \citep{PARSEC} isochrones to select models that have $\teff$, absolute Johnson $V$ magnitude and metallicity close to the observed ones for each star. Applying a Bayesian scheme described in \cite{2005A&A...436..127J} to selected models, Casagrande et al. derived masses and ages of stars. They used a flat prior on ages and a Salpeter initial mass function (IMF) as a prior for masses. 

\paragraph*{RAVE.}
There is a series of papers on distance estimations for stars in the RAVE survey \citep[DR5 is described in][]{2016arXiv160903210K}.  
\cite{2010A&A...511A..90B} proposed a method for distance estimation for RAVE stars based on a comparison of observed $\teff, \log g$, metal abundance $\met,$ and colour $(J-K_s)$ with $Y^2$ models \citep{2004ApJS..155..667D}. For each star 5000 realisations of observed parameters were sampled from a Gaussian distribution with dispersions equal to the measured uncertainties and for each realisation a closest model was selected. These authors took an average of the model parameters measured in all realisations to derive an absolute $J$ magnitude of the star $M_J$.  The difference between the derived absolute magnitude and visible $J$ magnitude from Two Micron All Sky Survey (2MASS) gives a distance. Extinction was ignored in this work. This approach is limited by the fact that it does not take into account the inhomogeneity of models in the $\teff - \log g$ plane, effectively increasing the weight for short evolutionary stages and decreasing it for longer ones. This issue was solved in \cite{2010A&A...522A..54Z} by weighting models with a weight proportional to age and mass range represented by each model. A likelihood depending on the difference between observed and model $\teff$ and $\log g$ was also added. Other important changes were applied, including a change from $Y^2$ to PARSEC \citep{PARSEC} isochrones, the addition of a prior on mass (assuming \cite{2003PASP..115..763C} IMF), and the application of a volume correction. Zwitter et al. calculated an absolute $J$ magnitude as a weighted mean of the absolute magnitudes derived from luminosities of the models. The difference between the visible $J$ magnitude from 2MASS and the absolute magnitude gives the distance modulus for each star. As in \cite{2010A&A...511A..90B}, extinction was ignored.

\cite{2014MNRAS.437..351B} further developed the above method by adding priors from the Galactic structure; they provide priors on age, metallicity, and positions from halo, thin, and thick disk models. A kinematic correction \citep[see][]{2012MNRAS.420.1281S} was also applied. Extinction was included into distance calculations. An exponential prior on the value of $ln(A_V)$ was imposed with the extinction value at infinity $ A_{V\infty}(b, l)$ taken from \cite{1998ApJ...500..525S}. The extinction at a given distance was calculated as $A_{Vprior}(b, l, s) = A_{V\infty}(b, l) \int_0^s \rho(s) ds / \int_0^\infty \rho(s) ds$, where $\rho(s)$ is the density of extincting material along the line of sight, taken from the model of the Galaxy \citep[see the Equation 10 in][]{2014MNRAS.437..351B}. This is so far the most advanced method and it was applied with minor modifications to LAMOST data as well (see below). Distance moduli (but not ages and masses) were recalculated with the same method for RAVE DR5.
\cite{2014MNRAS.437..351B} solved the problem of multimodal probability distribution functions (PDFs) for the distance modulus by fitting a Gaussian mixture model to it with up to three Gaussians. This approach works fine in most cases. However, as we illustrate below in \autoref{fig:pdf_example} it cannot be applied to mass and log(age) PDFs because they can be skewed or truncated, a shape which is hard to fit with a small set of Gaussians. Truncated shapes of the PDF arise from a limited range of allowed masses and log(age)s. For \cite{2014MNRAS.437..351B} limits are imposed by an age prior; see their Equations 3, 4, and 5.

\paragraph*{APOKASC.}
\cite{2014MNRAS.445.2758R} applied Bayesian methods to estimate distances and extinctions for approximately 2000 red giant stars from the joint APOGEE and Kepler Asteroseismic Science Consortium (APOKASC) sample \citep{2014ApJS..215...19P}, which is a part of APOGEE \citep{2014ApJS..211...17A}, covering the \textit{Kepler} field of view. They supplemented spectroscopic parameters $\teff$ and $\met$ with asteroseismic data from \textit{Kepler}. As alluded to before, from asteroseismic values $\Delta \nu$ and $\nu_{\rm{max}}$ and knowing $\teff$, it is possible to derive an estimate of stellar radius $R$ and mass $M$, using scaling relations from \citet{1995A&A...293...87K}, i.e.
\begin{eqnarray}
 \Delta \nu &\propto& M^{1/2} R^{-3/2} \\
 \nu_{\rm{max}} &\propto & M R^{-2} \teff^{-1/2}.
\end{eqnarray}
This puts more constraints on stellar models, thus increasing the precision of stellar parameters and distance determinations. Using PARSEC isochrones, \cite{2014MNRAS.445.2758R} built PDFs for stellar parameters (mass, radius, and surface gravity) and stellar absolute magnitudes. The latter were then combined with photometric data from  Sloan Digital Sky Survey (SDSS), 2MASS, and Wide-field Infrared Survey Explorer (WISE) to be converted to the PDFs of distance modulus $\mu_d$ and extinction $A_K$. The mode and 68\% confidence intervals of the PDFs were calculated for both distance and extinction. \cite{2014MNRAS.445.2758R} noted that over one-third of stars in their sample have bimodal PDFs. Bimodal PDFs were treated in the same way as single-peaked PDFs. Using the mode allows one to select the highest peak of the PDF and other peaks only show themselves by broadening of confidence intervals.
Only distance estimates were published by \cite{2014MNRAS.445.2758R}.

\paragraph*{LAMOST.}
The LAMOST team is also working on estimating distances to stars from spectroscopic data. First and second public data releases of the project include spectral properties for about one and two million stars, respectively \citep{2015RAA....15.1095L}. 

\cite{2015AJ....150....4C} used a Bayesian approach to derive distances from LAMOST DR1 data combined with 2MASS photometry. They used the Dartmouth Stellar Evolution Database \citep{2008ApJS..178...89D} and a Bayesian technique similar to that by \cite{2011A&A...532A.113B} to derive the PDF of the absolute magnitude for each star. This was then converted to distances using 2MASS photometry. Interstellar extinction was ignored in this work. 
\cite{2015AJ....150....4C} performed  a comparison with RAVE distances to test their method. The derived distances are systematically smaller by 12\% than those derived by \cite{2010A&A...522A..54Z} with 16\% spread. Given the precision of the LAMOST data, they derived distance uncertainties to be on the order of 40\%.

\cite{2016MNRAS.456..672W} applied the Bayesian approach from \cite{2014MNRAS.437..351B} to derive parallaxes and extinctions for LAMOST data. Again, 2MASS photometry was used. The reported uncertainty in parallax is about 20\% for dwarf stars and 40\% for giants. Kinematic correction \citep[see][]{2012MNRAS.420.1281S} was applied using PPMXL \citep{2010AJ....139.2440R} and UCAC4 \citep{2013AJ....145...44Z} data. Data from \cite{2015AJ....150....4C} and \cite{2016MNRAS.456..672W} are not yet publicly available. 

Distances are provided in the LAMOST Galactic Anti-Centre project data release \citep{2015MNRAS.448..855Y}. These data include spectroscopic measurements of \threeparams\ and photometry from 2MASS and Xuyi Schmidt Telescope Photometric Survey \citep{2014RAA....14..456Z}.
In their work, \cite{2015MNRAS.448..855Y} applied two different methods to get distances. In the first method, which they call ``empirical'', stars are divided into four groups (OB stars, giants, and two groups of dwarfs). For each group absolute magnitudes were calculated using a third-order polynomial of \threeparams. Polynomials were derived by fitting data from the parts of the medium resolution INT Library of Empirical Spectra (MILES) library \citep{2006MNRAS.371..703S} corresponding to each group. The precision of the obtained distance modulus is about $0.^m65$ for GKM giants and $0.^m3$ for other groups. A second, ``isochrone'' distance estimate was derived using the isochrones of Dartmouth Stellar Evolution Database \citep{2008ApJS..178...89D}. For each star a model with closest values of \threeparams\ was selected from the database. A difference between the visible magnitudes of the star and absolute magnitudes for the closest model provides distance. For both methods extinction values were derived from the LAMOST data using the star-pairs method, which is described in \cite{2014IAUS..298..240Y}. The ``isochrone'' method provides distances that are about 5 percent lower than those derived by the ``empirical'' method.

\paragraph*{}
Studies listed above use similar methods, but the implementation can vary, leading to different results even for the same input data. Moreover, while distances are typically calculated, mass and, most importantly,
age estimates are less common. The amount of complementary spectroscopic data available in different surveys calls for a more unified approach.
In this paper we present a Unified tool to estimate Distances, Ages, and Masses from spectrophotometric data (UniDAM). There are two major points in which we differ from studies listed above:

First, whereas most of the previously published studies were dedicated to data from a single survey, we processed data from several large surveys with one tool. 
For some surveys no data on distances, masses, and ages are publicly available to date. For others our results are consistent with previously published studies with the advantage that our catalogue was produced with the same method, isochrones, and priors on parameters for all surveys. Thus all differences in results for different surveys can be attributed to systematic differences in parameters determined in the spectroscopic surveys. We provide more details on spectroscopic surveys used in Section \ref{sec:catalog}. Another advantage of using many surveys simultaneously comes from the fact that different surveys probe different parts of the Galaxy because of different observing strategies and locations of telescopes. Therefore we do not simply increase the statistics, but have a more complete coverage of the Galaxy.

Second, we try to lift the degeneracy of the transformation from observed to stellar parameters by representing PDFs as sums of unimodal functions (unimodal sub-PDFs or USPDF) for each evolutionary stage. Thus we separate out physically different solutions. This allows us to increase the precision of stellar parameters for each solution.

\section{Data samples used}\label{sec:catalog}
We used observable parameters from a set of publicly available spectroscopic surveys in our work. 
All surveys were cross-matched with 2MASS \citep{2006AJ....131.1163S} and AllWISE \citep{2014yCat.2328....0C} to get the infrared photometry.
We used only ''clean`` photometry that is only bands that are not affected by low photometric quality, contamination, or confusion. This was achieved by taking only bands with 2MASS quality flag (\texttt{Qfl}) set to \texttt{'A'} and AllWISE bands with the contamination and confusion flag (\texttt{ccf}) set to zero and photometric quality flag (\texttt{qph}) set to \texttt{'A'}. We also requested that the reported uncertainty in magnitude has a positive value.
\autoref{tbl:catalog} summarises properties of the spectroscopic surveys from which we extracted our input data. We discuss some of them below, focusing on parameters for each survey, which we added or modified for our purposes. 

\begin{table*}
\begin{center}
\begin{tabular}{lrrd{3.1}ccc}
\toprule
Survey & N sources & Resolution & \MyHead{1.8cm}{$\median{\Delta T}$ (K)} & \MyHead{1.8cm}{$\median{\Delta \log g}\,$ (dex)} & \MyHead{1.8cm}{$\median{\Delta \feh}\,$ (dex)} & Reference \\ \midrule
APOGEE (DR12) & $88\,000$ & $22\,500$ & 91.5 & 0.11 & 0.03 & \prefcount \\
APOGEE (DR13)* & $89\,000$ & $22\,500$ & 91.5 & 0.11 & 0.03 & \prefcount \\
APOKASC & $2\,000$ & $22\,500$ & 91.5 & 0.11 & 0.03 & \prefcount \\
LAMOST-GAC (Main sample)* & $368\,000$ & $1\,800$ & 115 & 0.19 & 0.15 & \prefcount \\
LAMOST-GAC (Bright sample)* & $1\,075\,000$ & $1\,800$ & 100 & 0.15 & 0.13 & \prefcount \\
LAMOST-CANNON* & $450\,000$ & $1\,800$ & 96.3 & 0.13 & 0.05 & \prefcount \\ 
RAVE (DR5) & $450\,000$ & $7\,500$ & 92 & 0.20 & 0.10 & \prefcount \\
RAVE-on* & $450\,000$ & $7\,500$ & 85 & 0.14 & 0.07 & \prefcount \\
GCS* & $13\,800$ & $20\,000$ & 80 & 0.10 & 0.10 & \prefcount \\
SEGUE*   & $277\,500$ & $2\,000$ & 145 & 0.26 & 0.13 & \prefcount \\
Gaia-ESO (DR2)* & $7\,000$ & $16\,000$ & 50 & 0.10 & 0.07 & \prefcount \\
AMBRE* & $3\,400$ & $16\,000$ & 120 & 0.20 & 0.10 & \prefcount \\
GALAH (DR1)* & $28\,000$ & $10\,700$ & 108 & 0.30 & 0.11 & \prefcount \\
Mock & $4 \times 8\,000$ & - & 100 & 0.10 & 0.10 & - \\\bottomrule
\end{tabular}\caption{Spectroscopic surveys from which data were used (see Section \ref{sec:catalog} for more information on surveys).
First column contains an abbreviated designation of the survey with a star symbol (*) indicating surveys that were included in the final catalogue (other surveys are available as separate tables). The second column lists the approximate number of observations\ that were used in our work. The third column is the resolution of the spectra 
from which $\teff, \log g$ and $\feh$ were derived. 
The fourth, fifth, and sixth columns list median uncertainties in $\teff$, $\log g$ and $\feh$. The last column shows the reference.}
\tablebib{
\setcounter{refcount}{0}
\pprefcount \cite{2014ApJS..211...17A}; 
\pprefcount \cite{2016arXiv160802013S};
\pprefcount \cite{2014MNRAS.445.2758R}; 
\pprefcount \cite{2017arXiv170105409X}; 
\pprefcount \cite{2017arXiv170105409X}; 
\pprefcount \cite{2015ApJ...808...16N}; 
\pprefcount \cite{2013AJ....146..134K}; 
\pprefcount \cite{2016arXiv160902914C}; 
\pprefcount \cite{2011AA...530A.138C};
\pprefcount \cite{2009AJ....137.4377Y}; 
\pprefcount \cite{GAIA_ESO}; 
\pprefcount \cite{AMBRE}; 
\pprefcount \cite{GALAH}.
}
\label{tbl:catalog}
\end{center}
\end{table*}

\subsection{APOGEE and APOKASC}\label{sec:apogee}
We used APOGEE data from SDSS DR12 \citep{2015ApJS..219...12A} and DR13 \citep{2016arXiv160802013S}. We kept only those stars that belong to the Main Survey Targets\footnote{see APOGEE target selection description at \url{http://www.sdss.org/dr12/irspec/targets/}} and have their temperatures, gravities, and metallicities measured. Both DR12 and DR13 were used, as they differ mainly in spectroscopic calibration, and it is interesting to test how that influences the  estimates of age, mass, and distance. We include our results for DR13 data in our final catalogue, whereas results for DR12 data are provided as a separate table.

We use as a separate input survey the APOKASC sample \citep{2014ApJS..215...19P}, although it is in this context just a subset of APOGEE. Therefore the result for this sample is not included in our final catalogue. These data were used to compare the results of \cite{2014MNRAS.445.2758R} with the prospect of the inclusion of asteroseismic data (see Section \ref{sec:compare_apokasc}). 

\subsection{LAMOST}
The second public data release of the LAMOST project \citep{2015RAA....15.1095L} contains spectral parameters for over 2 million stars. However, the uncertainties in the stellar parameters reported, i.e. $170\,$K in $\teff$, $0.5\,$dex in $\log
g$ and $0.2\,$dex in $\feh$, are too high for these data to be used reliably for the model fitting. Therefore we decided not to use the main LAMOST dataset. We focused instead on the LAMOST Galactic Anti-Center (LAMOST-GAC) project second data release \citep{2017arXiv170105409X}. This data release contains spectral parameters for about one-third of a million stars in the direction of the Galactic anti-center in its main sample. The bright sample contains over a million stars from a larger area. A different processing pipeline was used by LAMOST-GAC team, which resulted in substantially lower parameter uncertainties of $115\,$K in $\teff$, $0.2\,$dex in $\log g,$ and $0.13\,$dex in $\feh$.

An additional dataset derived from LAMOST DR2 data was prepared with The Cannon tool \citep{2015ApJ...808...16N, 2016arXiv160200303H}. This tool allows the transfer of parameters from high-resolution APOGEE spectra to LAMOST data using stars observed by both surveys for the calibration. This method transfers APOGEE uncertainties in the measured parameters to the LAMOST data, improving the precision of obtained parameters. To account for calibration uncertainties we added in quadrature the median APOGEE absolute uncertainties to the formal uncertainties reported by \textit{The Cannon}. Another benefit of \textit{The Cannon} tool is that it measures the value of $[\alpha/\rm{Fe}]$, which is not provided by LAMOST. The \textit{The Cannon} tool was only calibrated for giant stars, which are available from APOGEE-LAMOST overlap and therefore the LAMOST-CANNON sample contains only giant stars.

\subsection{RAVE surveys}
The fifth data release of the RAVE project \citep{2016arXiv160903210K} contains spectral parameters for almost half a million stars. This release contains a flag indicating whether the fitting algorithm has converged, but it turns out that even for stars with this flag set to zero (indicating that the fit converged) there are clear concentrations of values of effective temperatures and gravities towards grid points. This feature is known to the community \citep[see][]{2014MNRAS.437..351B}.
We added in the output catalogue a flag that indicates if $\log \teff$ is within $0.01\,$dex of a grid point or if $\log g$ or $\feh$ are on a grid point. 
About one-third of the stars are affected by clustering around grid points. 

The RAVE-on \citep{2016arXiv160902914C} is a product of processing of original RAVE spectra with \textit{The Cannon} tool. Calibration set was constructed from the overlap of RAVE with APOGEE giants and K2/EPIC survey \cite{2016ApJS..224....2H}. In addition to \threeparams,\
the output of \textit{The Cannon} tool also contains the value of $[\alpha/\rm{Fe}]$ and abundances for several chemical elements.
The RAVE-on data refer to exactly the same stars as the main RAVE survey, but the reported stellar parameters might be slightly different and the quoted uncertainties are smaller, therefore we chose to use RAVE-on in our catalogue, providing results for the main RAVE survey in a separate table. 

\subsection{Geneva-Copenhagen survey}
Geneva-Copenhagen survey (GCS) is the only non-spectroscopic survey used in this work. GCS is a photometric survey, which contains \threeparams\ derived using Stromgren photometry.
We used GCS re-analysed data published by \cite{2011AA...530A.138C}. We exclude $15\%$ of the stars for which no estimates of \threeparams\  are provided or for which no photometry was present. The latter was mainly because a number of GCS sources are too bright for 2MASS.

\subsection{SEGUE}
We used SEGUE data from SDSS DR12 \citep{2009AJ....137.4377Y} with internal uncertainties from the SDSS database. We add in quadrature the internal and systematic uncertainties derived by \cite{2008AJ....136.2070A} of $130\,$K in $\teff$, 
$0.21\,$dex in $\log g$ and $0.11\,$dex in $\feh$. 
The SEGUE survey is based on SDSS photometry, which is deeper than 2MASS, therefore for about one-half of SEGUE targets no 2MASS or AllWISE photometry is available or the photometry is very uncertain. We do not use such stars in our work.

\subsection{Gaia-ESO}
For the Gaia-ESO survey, the data release 2 \citep{GAIA_ESO} was used. For nearly half of its nearly $15\,000$ spectra $\teff, \log g$ and $\feh$ are available. So we used approximately $7\,000$ sources from this survey. 

\subsection{AMBRE}
Atmospheric Parameters and Chemical Abundances from Stellar Spectra \citep[AMBRE;][]{AMBRE} project released parameters extracted from the automatic analysis of the ESO spectral data archives for over $4\,500$ observations (over $2\,000$ sources). No photometry or positional information are provided in the project data, so we attempted to get this information using target names. With the SIMBAD service \citep{2000A&AS..143....9W} we obtained positions for nearly $1\,500$ sources, having a total of $3\,400$ observations in the AMBRE survey.

\subsection{GALAH}
\citep{GALAH}  describe the GALactic Archaeology with HERMES (GALAH) survey first data release. 
Stellar parameters were derived for 2576 GALAH stars with the Spectroscopy Made Easy (SME) tool \citep{2012ascl.soft02013V}. These data were used as a training sample for \textit{The Cannon} tool, which was then used to derive stellar parameters for the rest of the survey. \citet{GALAH} provide typical uncertainties of \textit{The Cannon} tool used to derive spectral parameters and internal precision of the SME. We added them in quadrature to get  uncertainties of $108\,$K in $\teff$, $0.3\,$dex in $\log g$ and $0.11\,$dex in $\feh$.

\subsection{Mock survey}\label{sec:data_mock}
In addition to real survey data we also created a mock survey to test our UniDAM tool. In this case we have full control on both the input parameters for our tool and the desired output parameters of the star.
We produced mock surveys by sampling a number of models from PARSEC isochrones \citep{PARSEC} (see Section \ref{sec:iso}). 

We stress that the choice of models was aimed at covering model parameter space. So our mock survey does not resemble observed stellar surveys, which are typically magnitude-limited, nor a physical distribution of stars in masses and ages. We motivate our choice by the need to study the behaviour of our tool over a large parameter range. 
We chose isochrones with 8 different metallicities and 20 ages, which we selected at random. From each isochrone we randomly selected 20 models (with mass below 4 $\Msun$). We used $\teff, \log g, \feh$ as well as 2MASS and AllWISE magnitudes for each selected model. High-mass stars were excluded because of their rarity.
We took absolute magnitudes from PARSEC models as our ``observed'' magnitudes, thus setting the distance to 10 pc and extinction to zero. Parameter uncertainties were taken to be $100\,$K for $\teff$, $0.1\,$dex for $\log g$, and $\feh$, and $0^m.03$ for each magnitude $m_\lambda$, which is similar to uncertainties in real spectroscopic and photometric surveys.

We prepared four mock surveys. In the first survey we took spectral and photometric values as provided by the PARSEC models. In the second survey we perturbed photometric parameters with random Gaussian noise, while keeping original spectroscopic parameters. In the third we perturbed spectral parameters with random Gaussian noise, while keeping original photometry. In the last survey all parameters were perturbed.
Perturbation spread was always taken to be equal to the chosen parameter uncertainties. This allows us to control how uncertainties in observations influence our results.

\section{Isochrones}\label{sec:iso}
We used PARSEC 1.2S isochrones \citep{PARSEC}, which provide a large sample of models covering a wide range of stellar parameters. These data include effective temperatures, surface gravities, radii, and absolute photometric magnitudes for a models covering large ranges in metallicities, ages, and masses. We selected nearly three million models that cover the following ranges:
\begin{itemize}
  \item $10^{-4} : 0.06\,$dex in metallicity ($Z$), corresponding to $-2.2 : 0.6\,$dex in $\feh$
  \item $6.6 : 10.13$ in log(age) $\tau$, corresponding to $4\cdot 10^6 : 13.5\cdot10^9$ years  \item $0.09 : 67\,\Msun$ in mass
\end{itemize}

The density of models varies within the ranges indicated above. The reason for this is that isochrones are designed to reproduce details of stellar evolution. Therefore, there are a relatively large number of models covering some rapid stages of evolution and there are a lower number of models for stages of slow evolution. To account for this, we introduced a value $w_j$ that is a measure of the volume of the parameter space (metallicity, age, and mass) represented by each model. Otherwise we would be biased towards rare evolutionary stages. 

We calculated $w_j$ for each model as a product of width of the bin in each dimension represented by the model,
\begin{equation}
  w_j = w_{\rm{age}, j} w_{Z, j} w_{\rm{mass}, j}. \label{eq:weight}
\end{equation}
The PARSEC isochrone models are calculated for the bin mid-points and they are not equal to the average model in each bin, which makes the binning somewhat arbitrary.

The PARSEC isochrones are equally spaced in log(age)s $\tau$. Therefore the density of models with lower ages is higher than that of models with higher ages. This has to be compensated for to avoid a bias towards lower ages. We took for the age bin width $w_{\rm{age}}$ the time span represented by the isochrone. It is calculated as $w_{\rm{age}, j} = (10^{\tau_{j+1}} - 10^{\tau_{j-1}})/2$, so time span range is defined by mid-points between isochrones in age. 

Observations provide $\met$ or $\feh$, which are proportional to the logarithm of $Z$. We created a grid in $Z$, ranging from $10^{-4}$ to $0.05$, such that the spacing between the values of $\feh$ is smaller than the mean uncertainty $\sigma_{\feh}$ of iron abundance measure at a given $\feh$ in the most precise input data, i.e. APOGEE data. Typically, uncertainties in $\feh$ are smaller for metal-rich stars than for metal-poor stars with $\sigma_{\feh} \propto Z^{-0.15}$. Therefore $\Delta_Z$ -- the bin width in $Z$ -- is roughly $\Delta_Z \propto Z \sigma_{\feh} \propto Z^{1 - 0.15} = Z^{0.85}$. The width of the bin in $Z$ was used for $w_Z$, thus ensuring a flat prior in $Z$. 
To check the impact of this, we performed tests with $w'_Z$ proportional to the width of the bin in $\feh$. These tests showed that this difference has little impact on our results. 
This is caused by the fact that for given values of $\teff$ and $\log g$, stellar parameters like age, mass, and luminosity are changing slowly with $\feh$ and $Z$, so the variations in weights are second-order effects. Thus $w_Z$ is the second-order effect, but we kept it to keep the flat prior in physical quantity $Z$.

Masses for models were selected by the PARSEC algorithm to track the shape of the isochrone as well as possible. This results in more models in more curved parts of the isochrone. Such an approach produces heavily inhomogeneous coverage of the mass range, which has to be corrected for. We used for $w_{\rm{mass}}$ the width of the bin in mass.

One of nine evolutionary stages is assigned to every PARSEC model. We grouped these stages into main-sequence stars and giants ascending the red giant branch (pre-core-helium burning; stage I), core-helium burning stars (stage II), and asymptotic giant branch stars (post-core-helium burning; stage III). These stage labels were used to separate models with different internal structures.

\begin{table}
\begin{center}
\begin{tabular}{lp{4cm}l} \\ \toprule
Column & Description & Unit\\ \midrule
Z &  Metallicity & - \\
log(age/yr) &  Age & log(years) \\
M\_ini & Initial mass & $\Msun$ \\
M\_act & Actual mass & $\Msun$  \\
logL/Lo &  Luminosity & - \\
logTe &  Effective temperature & - \\
logG &  Gravity & -\\
... & Set of absolute magnitudes (see text) & mag \\
int\_IMF & Value of the cumulative IMF function $F(M_{ini})$ & - \\
stage &  Evolutionary stage & - \\ \bottomrule
\end{tabular}
\end{center}
\caption{PARSEC model columns (as named in the output of \url{http://stev.oapd.inaf.it/cgi-bin/cmd_2.7}).}\label{tbl:parsec}
\end{table}

For each model we used basic physical information (see \autoref{tbl:parsec}) and 2MASS and AllWISE absolute magnitudes that were derived from the luminosities. 
Other magnitudes are often available, but we did not use them for the reasons discussed below.

\section{Methodology}\label{sec:bayes}
The method used in our tool is similar to the Bayesian method described in \cite{2014MNRAS.445.2758R}. We introduced the vector $\bO$ for input (``observed'') parameters and their uncertainties $\bO = (\teff, \log g, \feh, m_\lambda, \sigma_{\teff}, \sigma_{\log g}, \sigma_{\feh}, \sigma_{m_\lambda})$. Here, $m_\lambda$ indicates visible magnitudes in several photometric bands and $\sigma_x$ is the uncertainty of the parameter $x$. These values were taken from surveys listed in \autoref{tbl:catalog}. When $\alpha$ element abundances were available, the metallicity $\feh$ was corrected with the relation $\feh = \feh_0 + \log(1. + 0.638[\alpha/\rm{M}])$ \citep[see][]{1993ApJ...414..580S}. Additional input parameters for each star can be used in $\bO$, for example masses and radii derived from asteroseismic data or parallaxes from Gaia. 

We used two vectors for output parameters. The first vector, $\textbf{X}_m = (\tau, M, \feh)$ represents stellar model parameters log(age), mass, and metallicity. These parameters are taken from isochrone models and therefore have discrete values. We always refer to the actual rather than initial stellar mass because this quantity can be measured from other data, for example from asteroseismic quantities or from binary orbital solutions. The second vector, $\bX_p = (\mu_d, A_K),$ where subscript $p$ stands for photometry, represents distance modulus and extinction; these parameters can formally have any value, but we set a physically motivated limit $A_K \geq 0$ (see discussion in \autoref{sec:cuts}). 
The full output parameter vector is then $\bX = \bX_m \cup \bX_p$.

The probability of having parameters $\bX$ with given observables $\bO$ can be expressed via Bayesian formula as
\begin{equation}
 P(\bX|\bO) = \frac{P(\bX) P(\bO|\bX)}{P(\bO)} \propto P(\bX) P(\bO|\bX).\label{eq:bayes}
\end{equation} 

We used flat priors on age (in linear scale) and metallicity $Z$, which means a star formation rate that is constant in time and is independent of $Z$. The quantitative effect of different priors is described below in Section \ref{sec:compare_rave}.
We used a mass prior based on the  IMF $F_\textrm{IMF}$ from \citet{2003ApJ...598.1076K}. Therefore,
\begin{equation}
 P(\bX) = F_\textrm{IMF}(\mass).\label{eq:prior}
\end{equation}

Isochrones give us for each $\bX_m$ a new vector $\textbf{O'} = (\teff, \log g, \feh, M_\lambda)$, where $M_\lambda$ indicates absolute magnitudes in several photometric bands. So we can define a function $\mathcal{I}$ as $\textbf{O'} = \mathcal{I}(\bX_m)$. Noticeably, $\feh$ is contained in both $\bO$ and $\bX_m$, so $\mathcal{I}_{\feh}(\bX_m) \equiv \feh$. 
We express $P(\bO|\bX)$ using two log-likelihoods
\begin{equation}
 P(\bO|\bX) = P(\bO|\bX_m, \bX_p) = e^{-L_{iso}-L_{sed}}.
\end{equation} 

Here, $L_{iso}$ is a measure of the separation between observed spectral parameters \threeparams\ and those predicted by model parameters $\bX_m$. Assuming Gaussian uncertainties in $\bO$, we can write
\begin{align}
 L_{iso} &= \sum_{i \in (\teff, \log g, \feh)} \frac{(O'_i - O_i)^2}{2 \sigma^2_{O, i}} \nonumber \\
         &= \sum_{i \in (\teff, \log g, \feh)} \frac{(\mathcal{I}_i(\bX) - O_i)^2}{2 \sigma^2_{O, i}}\label{eq:liso}.
\end{align}

We use \autoref{eq:liso} for the log-likelihood as in most cases spectroscopic surveys do not provide information about the correlations of the uncertainties on the different parameters.
When this information or the PDFs of the spectroscopic parameters are known, this information can be included in\ $L_{iso}$.

$L_{sed}$ is a measure of similarity of the observed spectral energy distribution (SED) (set of $m_\lambda$ obtained from photometric surveys) and that predicted by isochrone model for $\bX$. Visible magnitudes $m_\lambda$  come from 2MASS and AllWISE. These magnitudes are related to the absolute magnitudes $M_\lambda$ in $\bO'$ by the following relation: 
\begin{equation}
 m_\lambda = \mu_d + M_\lambda + A_\lambda = \mu_d + M_\lambda + C_\lambda A_K,
\end{equation} 
where the extinction in band $\lambda$ is defined as $A_\lambda = \frac{R_\lambda}{R_K} A_K = C_\lambda A_K$, with the extinction coefficients $R_\lambda$ taken from \cite{Extinction} and summarised in \autoref{tbl:rlambda}. 
Therefore to compare observed visible and model absolute magnitudes we need to know the distance modulus $\mu_d$ and the extinction $A_K$ in the direction of the star. The latter can be obtained from extinction maps or by comparing magnitudes in different bands.
We chose the second method and calculated the extinction value for each star using photometric infrared data. This allowed us to take into account variations of extinction that might occur on scales smaller than the typical map resolution. 
We were also able to calculate extinction for any position on the sky and any distance, whereas a detailed three-dimensional extinction maps are created only for the nearest kiloparsec \citep{2012AstL...38...87G} or for the Galactic plane \citep{2014MNRAS.443.2907S}, which is not sufficient for our purpose. A more recent three-dimensional map by \citet{2015ApJ...810...25G} covers a large fraction of the sky, but a full-sky three-dimensional extinction map is still not available.

We use 2MASS and AllWISE data as infrared bands are much less affected by interstellar extinction. Extinction in optical bands is generally higher than in the infrared and can have higher spectral variations between different points on the sky \citep[see a discussion in ][]{2011ApJ...739...25M}. By using infrared data alone we increased the precision of our distance estimates at a cost of decreasing the precision for the extinction estimate. As far as we focus on distances, this seems a fair trade.

\begin{table}
\begin{center}
\begin{tabular}{ccc} \\ \toprule
Band & Value of $R_\lambda$ & Value of $C_\lambda$ \\ \midrule
J & 0.720 & 2.35\\
H & 0.460 &1.5\\
K & 0.306 & 1 \\
W1 & 0.180 & 0.59 \\
W2 & 0.160 & 0.52 \\ \bottomrule
\end{tabular}
\end{center}
\caption{Values of the extinction coefficients $R_\lambda$ and $C_\lambda$ used for 2MASS and AllWISE photometry. Values were taken from \citet{Extinction}}\label{tbl:rlambda}
\end{table}

For $L_{sed}$ we use the following expression:
\begin{equation}
  L_{sed} = \sum_\lambda \frac{(m_\lambda - M_\lambda - C_\lambda A_K - \mu_d)^2}{2 \sigma_{m_\lambda}^2} - V_{corr}(\mu_d),
\end{equation}
 where the summation is carried out over all bands, for which photometry is available for a given star and  $V_{corr}(\mu_d)$ is a volume correction.
 We introduce volume correction to compensate for the fact that with a given field of view we probe larger space volume at larger distances than at smaller distances. See a discussion of the effect of volume correction in \autoref{sec:priors}. Using the relation between distance modulus and distance ($d = 10^{0.2\mu_d + 1}$), we can write
\begin{equation}
 V_{corr}(\mu_d) = \log d^2 = \log 10^{2 (0.2 \mu_d + 1)} = (0.4 \mu_d + 2) \log 10.
\end{equation} 

We use both $L_{iso}$ and $L_{sed}$ in $P(\bO|\bX)$. Therefore on top of the spectroscopic parameters we utilise
additional information, namely, the SED of the star. The drawback is that this also brings in the systematic errors of both stellar spectra modelling in PARSEC and possible large errors in photometry in the case of a mismatch between spectroscopic and photometric surveys.

\subsection{Probability distribution functions}\label{sec:pdf}

In order to get the PDF in each parameter, we need to marginalise a multi-dimensional PDF of output parameters, $P(\bX|\bO)$, defined in \autoref{eq:bayes}, over all other parameters. For example, for log(age) $\tau$ one has to calculate
\begin{equation}
 P(\tau) = \iiiint P(\bX) P(\bO|\bX) dM d\feh d\bX_p,
\end{equation} 
with the integral taken over the whole parameter space.
In practice, we have a discrete sample of models from isochrones. So we can replace $P(\bO|\bX)$ with the sum of delta functions 
\begin{equation}\
   P(\bO|\bX) = \sum_j P(\bO|\bX_{m,j}, \bX_p) \dtau \dM \dfeh w_j,\label{eq:pox}
\end{equation}
where we write for brevity $\dtau = \delta(\tau_j - \tau)$, $\dfeh = \delta(\feh_j - \feh)$, $\dM = \delta(M_j - M)$. Here we have to use volumes represented by each model $w_j$ from \autoref{sec:iso}, which reflect the volume of the parameter space represented by the model. The summation is carried out over all models and $\bX_{m,j} = (\tau_j, M_j, \feh_j)$ is a vector of parameters of the model $j$. 
Therefore we can write, using equations \ref{eq:prior} and \ref{eq:pox},
\begin{align}
 P(\tau) = \sum_j \iiiint & F_{IMF}(M) P(\bO|\bX_{m,j}, \bX_p) \dtau \dM \dfeh \nonumber \\ 
          & \times w_j d M\,d\feh\,d\bX_p, \label{eq:p_tau}
\end{align}
with again the summation carried out over all models.  We need to keep integration over $\bX_p$, because both $\mu_d$ and $A_K$ are continuous values.

We can make two important simplifications here.
First, there is no need to sum over all models because for most of them $P(\bO|\bX)$ is very small. We chose a threshold of $L_{iso} < 8$. In this case \threeparams\ are within 4 sigma uncertainties from observed values. We verified that increasing the threshold from 8 to 12.5 (or going from a combined 4 sigma to 5 sigma uncertainty threshold) leads to marginal changes in the output; parameter estimates change by more than 3\% for only less than 2 percent of the stars. Because models are selected from three-dimensional space, this comes at a cost of doubling the number of models to be considered. A decreasing the likelihood threshold however leads to more significant changes in the resulting parameters.

Second,  $L_{sed}$ is a quadratic form in $\mu_d$ and $A_K$. Therefore, for a given model $j$, $P(\bO|\bX_{p,j}) = \exp(-L_{sed})$ is a bivariate Gaussian distribution. The location of the maximum of this distribution can be found by solving the system of equations as follows:
\begin{equation}
\large{
 \begin{cases}
   \deriv{L_{sed}}{\mu_d} &= 0 \\
   \deriv{L_{sed}}{A_K} &= 0
 \end{cases}}.
\end{equation}
This is equivalent to the following set of equations:
\begin{equation}\label{eq:old_system}
\large{
\begin{cases}
  \sum_\lambda \fracdelta{1}\mu_d + \sum_\lambda \fracdelta{C_\lambda} A_K &= \sum_\lambda \fracdelta{m_\lambda - M_\lambda} - 0.4 \log 10 \\
  \sum_\lambda \fracdelta{C_\lambda} \mu_d + \sum_\lambda \fracdelta{C_\lambda^2} A_K &= \sum_\lambda \fracdelta{C_\lambda(m_\lambda - M_\lambda)},
\end{cases}}
\end{equation}
which is solved for $\mu_d$ and $A_K$. If $A_K < 0$, which is not physical, or if only one magnitude is available we set $A_K = 0$ and obtained $\mu_d$ from the first part of \autoref{eq:system}. By doing so we increase $L_{sed}$ for a given model, which decreases the contribution of this model to the PDFs. In some cases $A_K$ is zero for all models for a star. This indicates either that the extinction for this star is statistically indistinguishable from zero or that there is a mismatch between spectral and photometric data, which results in using visible magnitudes from a different star.
In the first case we still produce a reliable distance estimate, while in the second case the obtained $L_{sed}$ is high and the quality of the result is low (see Section \ref{sec:p_best_p_sed}).

The covariance matrix of $P(\bO|\bX_{p,j})$ is exactly the inverse Hessian matrix $H$ of $L_{sed}$ , i.e.
 \begin{equation}
 \large{
  H = 
    \begin{vmatrix}
     \frac{\partial^2 L_{sed}}{\partial \mu_d^2}            & \frac{\partial^2 L_{sed}}{\partial \mu_d \partial A_K} \\
     \frac{\partial^2 L_{sed}}{\partial \mu_d \partial A_K} & \frac{\partial^2 L_{sed}}{\partial A_K^2}
    \end{vmatrix} =
    \begin{vmatrix}
     \sum_\lambda\fracdelta{1}         & \sum_\lambda\fracdelta{C_\lambda} \\
     \sum_\lambda\fracdelta{C_\lambda} & \sum_\lambda\fracdelta{C_\lambda^2}
    \end{vmatrix}
    }.
 \end{equation}
It is important to note that $H$ depends only on $C_\lambda$, which are constants and photometric uncertainties $\sigma_{m_\lambda}$, thus $H$ has a constant value for a given star.  

The width of the $L_{sed}$ distribution in $\mu_d$ for a given model is thus $\Delta_{\mu_d} = \sqrt{H^{-1}_{0,0}}$,
which is of the order of $\sigma_m$ and is about an order of magnitude smaller than a typical  uncertainties in $\mu_d$ that we derive. This is not true for the extinction, but we are not focused on derivation of high-quality extinctions. Moreover, tests show that the error we bring into mean extinction values by this simplification is small. Furthermore  there is an obvious correlation between $\mu_d$ and $A_K$, but we ignore it here. This is justified because $\Delta_{\mu_d}$ is approximately one order of magnitude smaller than a typical  uncertainties in $\mu_d$ for a star, so a correlation between derived $\mu_d$ and $A_K$ in a two-dimensional PDF $P(\mu_d, A_K)$ for a given star is dominated by scatter in $\mu_d$ and $A_K$ for models used to build the PDF, rather than by correlation of $\mu_d$ and $A_K$ for each model.

As far as $P(\bO|\bX_p)$ is a bivariate Gaussian function, the integral over $d\bX_p$ in \autoref{eq:p_tau} is exactly the value of $P(\bO|\bX_p)$ at the location of its maximum, derived in \autoref{eq:system}. So we can replace the integral in \autoref{eq:p_tau} with a delta function
\begin{align}
 P(\tau) &= \sum_j \iiiint F_{IMF}(M) P(\bO|\bX) \dtau \dM \dfeh \nonumber \\ 
        & \times  \delta(\bX_{p,j} - \bX_p) w_j d M d \feh d \bX_p \label{eq:pdf} \\ 
        & = \sum_j P(\bO|\bX_{m,j}, \bX_{p,j}) F_{IMF}(M_j) \delta(\tau_j - \tau) w_j, \nonumber 
\end{align}
where $\bX_{p,j}$ is the solution of \autoref{eq:system} for model $j$. A similar equation can be written for $P(M)$.

For $\mu_d$ and $A_K$ each model contributes to the PDF a Gaussian summand with a width of $\Delta_{\mu_d}$ and $\Delta_{A_K} = \sqrt{H^{-1}_{1,1}}$, respectively. We can correct for using delta function in place of a bivariate Gaussian for $P(\bO|\bX_p)$ by adding a Gaussian smoothing multiplier with the corresponding width
\begin{subequations}
\begin{eqnarray}
  P(\mu_d) &= \sum_j P(\bO|\bX_{m,j}, \bX_{p,j}) F_{IMF}(M_j) w_j e^{-\frac{(\mu_{d,j} - \mu_d)^2}{2 \Delta_{\mu_d}^2}} \\
  P(A_K)   &= \sum_j P(\bO|\bX_{m,j}, \bX_{p,j}) F_{IMF}(M_j) w_j e^{-\frac{(A_{K,j} - A_K)^2}{2 \Delta_{A_K}^2}}.
\end{eqnarray}\label{eq:pdf_mu}
\end{subequations}

\subsection{Quality of model fit to the data}\label{sec:p_best_p_sed}
It is important to quantify how well a set of models represents the observed parameters of a star. To accomplish this, we used the $\chi^2$ distribution to get the $p$ values from our log-likelihoods. 
We characterise our model quality by the $p$ value corresponding to the value of $\chi^2 = L_{iso}+L'_{sed}$ for the model with the highest $P(\bO|\bX_{m,j}, \bX_{p,j})$. Here we use $L'_{sed} = L_{sed} + V_{corr}(\mu_d)$. We added $V_{corr}(\mu_d)$, thus removing the volume correction. This is necessary because $V_{corr}(\mu_d)$ adds a non-$\chi^2$ summand, that depends on $\mu_d$. Thus, the $\chi^2$ value used to compute the $p$ value is not the lowest possible value, but that corresponding to the model with the highest $P(\bO|\bX_{m,j}, \bX_{p,j})$, thus the value that is closest to observables. 
The number of degrees of freedom in this case equals the number of observables, which is the number of available magnitudes plus three (for the temperature, surface gravity, and metallicity dimensions). This $p$ value is designated as $p_{best}$. Low values of $p_{best}$ can be caused either by observables falling out of the range covered by the models or by inconsistencies between observed stellar parameters and the observed SED. We flagged data with $p_{best} < 0.1$ (see Section \ref{sec:quality}).

In addition to $p_{best}$, we quantify how good our models are at representing the observed SED. We use the same model used for $p_{best}$ and report as $p_{sed}$ the $p$ value corresponding to the chi-square value $\chi^2 = -L'_{sed}$. The number of degrees of freedom in this case is equal to the number of available magnitudes. Low values of $p_{sed}$ might be caused, for example, by a mismatch between spectral and photometric data, which results in using visible magnitudes from a different star. Another possible reason is a problematic spectral parameter estimation, which makes $\teff$ inconsistent with the SED from photometry. We flagged data with $p_{sed} < 0.1$ (see Section \ref{sec:quality}).

\subsection{Calculating final values}\label{sec:final_values}
From \autoref{eq:pdf} we can define a weight for each model $j$ as
\begin{equation}\label{eq:model_weight}
  W_j = P(\bO|\bX_j)F_{IMF}(M_j) w_j = e^{-L_{iso}-L_{sed}} w_j F_{IMF}(M_j).
\end{equation}

The PDF in each parameter can thus be calculated as a distribution of parameters for models, with weights $W_j$. For the PDF in $\mu_d$ and $A_K$ we smooth histograms with Gaussian kernel of width $\Delta_{\mu_d}$ or $\Delta_A$, respectively (see \autoref{eq:pdf_mu}).

\subsubsection{Determination of unimodal sub-PDFs}
\begin{figure*}
 \myimage{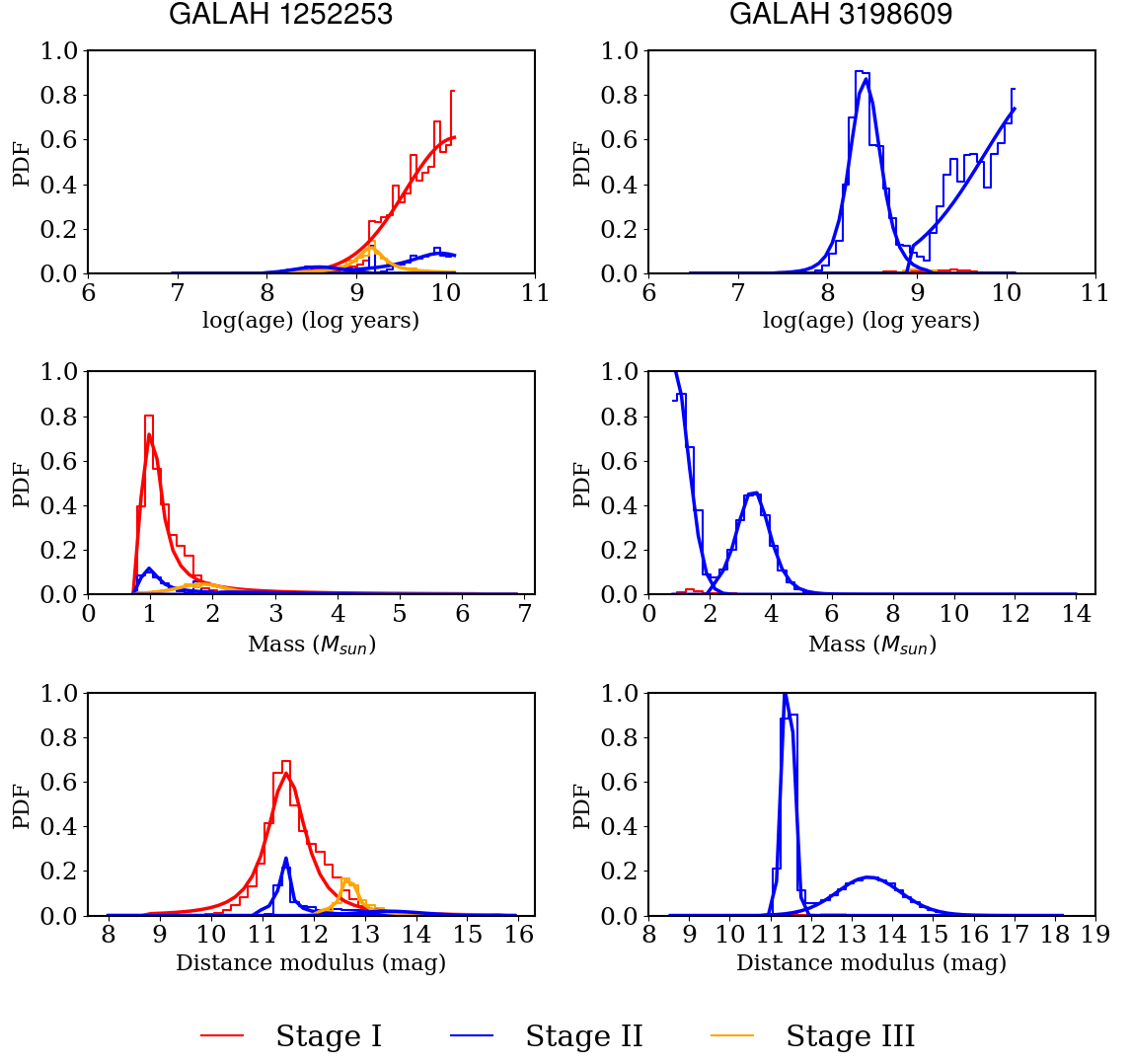}
 \caption{Probability distribution function for log(age) (top row), mass (middle row), and distance modulus (bottom row) for two stars from our catalogue. Different evolutionary stages are indicated with different colours (see legend). The stages are main-sequence stars and giants ascending the red giant branch (pre-core-helium burning, stage I), core-helium burning stars (stage II) and asymptotic giant branch stars (post-core-helium burning, stage III).
 The histograms represent model distributions and solid lines indicate our fits for individual USPDFs.} \label{fig:pdf_example}
\end{figure*}

For each combination of stellar mass, age, and metallicity, PARSEC models provide a single combination of effective temperature and surface gravity. The transition from the effective temperature, surface gravity, and metallicity to stellar mass and age is however non-unique. For a given combination of \threeparams\ with their uncertainties it is possible to find more than one corresponding model with different combinations of age and mass. For example, red clump stars and red giant stars can have similar spectral parameters \threeparams, but different ages, masses, and internal structures. Therefore, distributions in ages, masses, and absolute magnitudes (and thus distances) are in some cases different from Gaussian. This is illustrated in \autoref{fig:pdf_example}, where we show typical PDFs in log(age), mass, and distance modulus for two stars. Some of the distributions shown are multimodal with two or more peaks. Reporting mean values and standard deviations do not capture that properly. Mode values, as used by \cite{2014MNRAS.445.2758R}, give the value of the highest peak only. Full distributions can be provided, but they are often considered too complex for further analysis.
We suggest an intermediate solution: split PDFs into several USPDFs with each of these described by a unimodal function, assuming that this represents a group of models with similar stellar structure.

We split all models in three evolutionary stages, described in Section \ref{sec:iso}, i.e. pre-core-helium burning, core-helium burning, and post-core-helium burning stars (plotted in \autoref{fig:pdf_example} with red, blue, and yellow, respectively). Splitting our results this way has a benefit in case of overlapping isochrones from different evolutionary stages for a given $\teff, \log g, \feh$ combination. Without this split, we would combine values for substantially different evolutionary stages that are not physical. 

Splitting in evolutionary stages is not enough, as due to curvature of isochrones we can have sub-groups of models within one stage that are physically different. A good example is stage I, which contains both main-sequence and giant stars. This results in multimodal distributions of models in space of stellar parameters. An example of such a situation is given in \cite{2005A&A...436..127J} (see their Figures 1 and 2).
To split a multimodal distribution into several unimodal distributions we applied an additional empirically derived routine to our PDFs, which is described in \autoref{sec:split_uspdf}.
This routine works in the vast majority of cases. Those cases in which our splitting of the PDFs breaks down typically have too few models to produce a histogram (such cases are given the quality flag ``N'', see Section \ref{sec:quality}). 

The overall weight $V_m$ for a USPDF $m$ is defined as a ratio of the sum of weights of models within the USPDF and the sum of weights of all models
\begin{equation}
  V_m = \frac{\sum_{j \in m}W_j}{\sum W_j}.
\end{equation}

\subsubsection{Output values}\label{sec:outvalues}
We provide output for each USPDF such that it is possible to reproduce the PDF in each parameter we are interested in mass $\mass$, logarithm of age $\tau$, distance modulus $\mu_d$, distance $d$, parallax $\pi$, and extinction $A_K$.
Even for a unimodal distribution the mean, median, or mode might be a poor estimate for the value of interest in case the distribution is non-symmetric. Values of mode and median might produce less bias but should be used with care as they are not proper moments of the PDF and many statistical methods rely on moments.
To provide a simple representation of each USPDF (for each parameter), we fit them with a Gaussian, a skewed Gaussian, a truncated Gaussian, a modified truncated exponential distribution (MTED) and a truncated \stud\ (see definitions of these functions in the Appendix \ref{app:functions}). For the truncated functions the upper and lower truncation limits were not fitted, but were set to upper and lower limits of the considered USPDF. We selected the function that gives the lowest symmetric Kullback–Leibler divergence value, which is the measure of the information gain
\begin{equation}
  D_{KL} = \sum_i H_i \log \frac{H_i}{F_i},
\end{equation}
where $H_i$ are histogram counts and $F_i$ are fitted function values.

Truncated functions were taken because we have a natural upper limit on the age of the star, which is the age of the Universe and therefore there is a lower limit on the mass of a star that left the main sequence, which is approximately $0.7 \Msun$ if we consider the full range of metallicities. This limit produces sharp cut-offs in histograms, making truncated functions a natural choice. 

A modified truncated exponential distribution in log-age is equivalent to a flat distribution in ages, thus such a fit indicates that age is poorly constrained. 
Such PDFs are typical for main-sequence stars, where, as expected, it is hard to constrain age from spectrophotometric data.

In rare (less than $1\%$) cases in which the fit did not converge for all five fitting functions. This is primarily caused by long tails in the distributions or insufficient data for a proper fit. 
In this case we reported only mean and standard deviation of the data as fit parameters.

\begin{figure*}
 \myimages{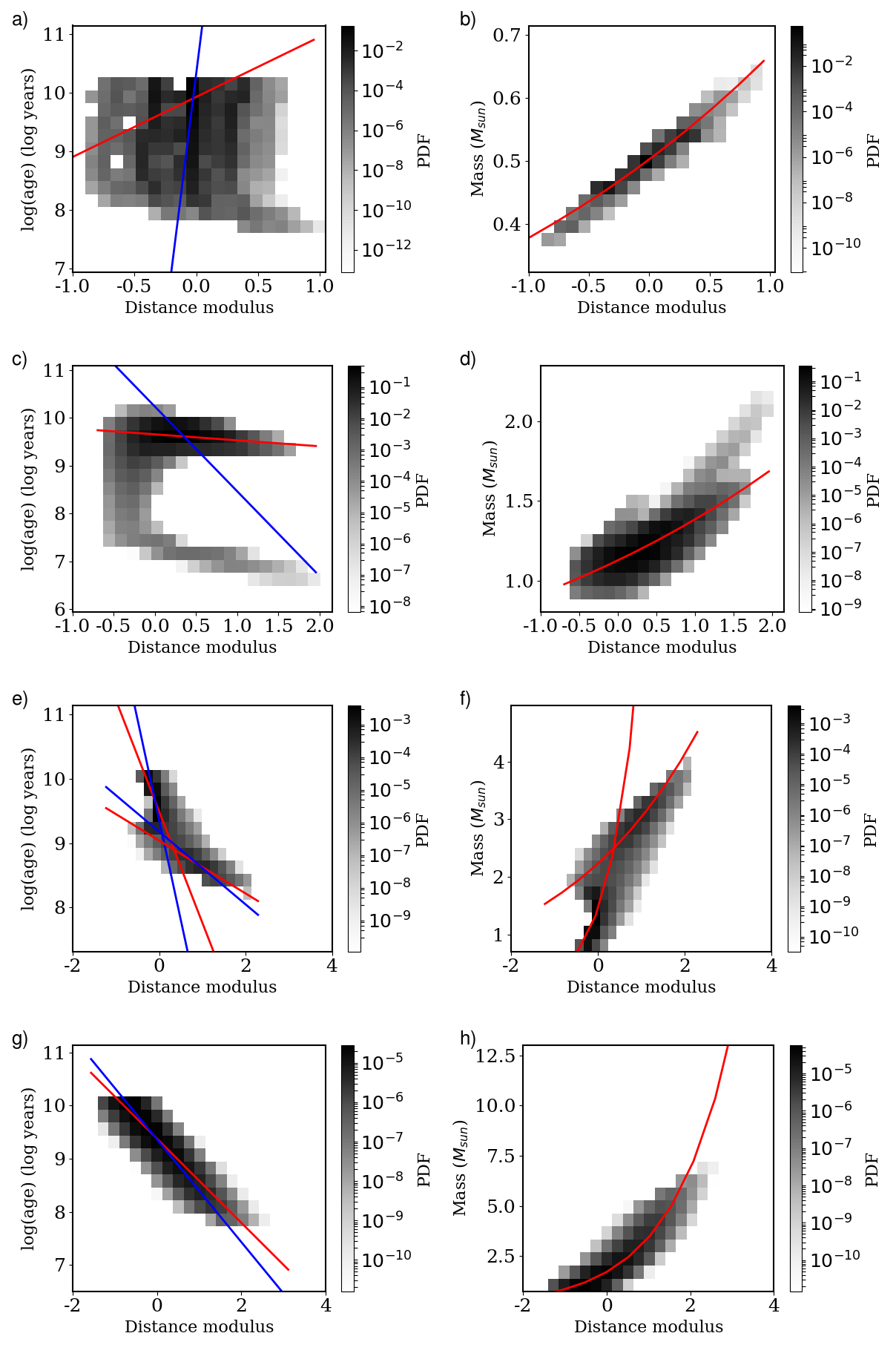}{0.8}
 \caption{Two-dimensional PDFs for typical lower main-sequence (a and b), lower giant branch (c and d), red clump (e and f), and upper giant branch (g and h) stars. The left column shows distance modulus and log(age) PDFs and the right column shows distance modulus and mass PDFs. The red line shows $\tau(\mu_d)$ and $\log M(\mu_d)$ fits and the blue line shows - $\mu_d(\tau)$ fit. Shading indicates PDF values (in log scale). Panels e and f have two lines of each colour because we detected two USPDFs for this star.}\label{fig:2d}
\end{figure*}

An important property of our result is that values of distance, mass, and log(age) for models are strongly correlated. We used the fact that distance modulus, log(age), and logarithm of mass have a nearly linear correlation within every USPDF in most cases. 
We report coefficients (slope $a$ and intercept $b$) of a weighted linear fit and a scatter around it for three relations for each USPDF:
$\tau = a_1 \mu_d + b_1$ (distance modulus versus logarithm of age; red lines on left panels of \autoref{fig:2d}),
$\mu_d = a_2 \tau + b_2$ (logarithm of age versus distance modulus; blue lines on left panels of \autoref{fig:2d}), and
$M = 10^{a_3 \mu_d + b_3}$ (distance modulus versus logarithm of mass; red lines on right panels of \autoref{fig:2d}). An illustration of correlations is given in \autoref{fig:2d}, where we show the two-dimensional PDFs for several stars and our fits to them. From the right-hand side panels it is clear that the relation of distance modulus and logarithm of mass is close to linear; the mass is plotted in linear scale, thus our fits are not straight lines. The relation of distance modulus and log(age) is weak for the main-sequence stars and lower giant branch stars. The shape of two-dimensional PDF can be quite complex, like in panels a and c of \autoref{fig:2d}. For these cases the scatter is large and our relations does not work.
For giant stars these correlations are much more pronounced, as can be seen in panels e and g of \autoref{fig:2d}. 
Correlations between distance modulus, log(age), and log(mass) can be used, for example, if the new distance estimate $\mu'_d$ is obtained from some external source (like Gaia) for a star in our catalogue. We verified that our estimates for mass and log(age) can then be corrected for by the value of the slope times the difference between the externally determined distance modulus and our estimate as follows:\ 
\begin{eqnarray}
 \tau' &=& \tau + a_1 (\mu'_d - \mu_d) \\
 M' &=& M \times 10^{a_3 (\mu'_d - \mu_d)}.
\end{eqnarray}
In the future work we will show applications of these relations, which are beyond the scope of this work.

Summing up, we chose to provide for each USPDF $m$ and each stellar parameter designated as $Y_{i, m}$, where $Y_i \in (\mass, \tau, \mu_d, d, \pi, A_K)$ the following quantities:
\begin{itemize}
\item A weighted mean (catalogue column suffix \texttt{\_mean}) of model values $Y_{i,j}$ for all models,
\begin{equation}
  Y_{i,m} = \frac{\sum_{j \in m} W_j Y_{i,j}}{\sum_{j \in m}W_j},
\end{equation}
where the summation is carried out over all models within USPDF $m.$
\item A weighted standard deviation (suffix \texttt{\_err}),  
\begin{equation}
  \sigma_{Y_{i, m}} = \sqrt{\frac{\sum_{j \in m} W_j (Y_{i,j} - Y_{i, m})^2}{\sum_{j \in m}W_j}}
.\end{equation}
\item A mode of the USPDF (suffix \texttt{\_mode}).
\item A weighted median value (suffix \texttt{\_median}).
\item A character indicating which fitting function was chosen: <<G>> for Gaussian, <<S>> for skewed Gaussian, <<T>> for truncated Gaussian, <<L>> for MTED, <<P>> for truncated \stud, <<E>> if the fit failed for all five functions, and <<N>> if there was not enough data for a fit (suffix \texttt{\_fit}). 
\item Parameters for a chosen fit (suffix \texttt{\_par}). The first two values are location and shape for the chosen best fitting function. For a Gaussian function, by definition, location parameter is equal to the mean value and shape parameter to the variance. If the chosen function is a skewed Gaussian then the third value is the skew value. If the chosen function is a truncated Gaussian or MTED, then third and fourth values are lower and upper limits. If the chosen function is \stud\ then the third value is the number of degrees of freedom and the fourth and fifth values are lower and upper limits.
\item One- and three- sigma confidence intervals. These are defined as a region containing 68.27\% (for one-sigma uncertainties) or 99.73\% (for three-sigma uncertainties) of the USPDF, positioned to minimise its span. By construction, such a confidence interval always includes the mode value. For a Gaussian distribution this is equivalent to a range centred on the mean value with width of one- or three- standard deviations. (suffixes \texttt{\_low\_1sigma, \_up\_1sigma, \_low\_3sigma, \_up\_3sigma}).
\end{itemize}

We report all USPDFs with weights $V_m$ higher than $0.03$. Integer priority values starting from 0 were assigned to each USPDF in order of decreasing weights $V_m$. We list all measures provided for our catalogue in \autoref{tbl:output}.

In \autoref{fig:pdf_example} we show examples of different USPDFs and fits to them. For the first star (left column) three different evolutionary stages are possible. For the evolutionary stage I a truncated Gaussian is required to fit log(age) distribution and for the evolutionary stage II a skewed Gaussian is needed to fit distribution in mass and distance modulus. For the second star (right column) mainly stage II is possible, but the distribution for this stage can be split in two parts.   We need a truncated \stud\ to fit the histogram for the higher age solution. The small USPDF visible for mass around $1 \Msun$ and log(age) of $9.3$ for stage I was excluded because its  weight $V_m$ is below the accepted $0.03$ threshold.

\begin{table}
\begin{center}
\begin{tabular}{lcp{4cm}} \\ \toprule
Column name & Units & Description \\ \midrule
id    &   & Unique ID of the star from the input data \\
stage &   & Stage number (I, II or III) \\
uspdf\_priority & & Priority order of a given USPDF (starting from 0) \\
uspdf\_weight & & Weight $V_m$ of a given USPDF \\
total\_uspdfs & & Number of USPDF with $V_m > 0.03$ \\
p\_best & & Probability for a best-fitting model (see Section \ref{sec:p_best_p_sed}) \\ 
p\_sed & & $p$-value from $\chi^2$ SED fit (see Section \ref{sec:p_best_p_sed})\\ 
quality & & Quality flag (see Section \ref{sec:quality}) \\
distance\_modulus$\dagger$ & mag & Distance modulus $\mu_d$ \\
distance$\dagger$  & kpc & Distance $d$ \\ 
parallax$\dagger$ & mas & Parallax $\pi$\\
extinction$\dagger$ & mag & Extinction $A_K$ in 2MASS K-band \\
mass$\dagger$ & $\Msun$ & Mass \\
age$\dagger$ & log(yr) & Logarithm of age $\tau$\\ 
\multicolumn{3}{c}{Distance modulus - logarithm of age relation:} \\ 
dm\_age\_slope & & Slope of the relation \\
dm\_age\_intercept & & Intercept of the relation \\
dm\_age\_scatter & & Scatter of the relation \\
\multicolumn{3}{c}{Logarithm of age - Distance modulus relation:} \\ 
age\_dm\_slope & & Slope of the relation \\
age\_dm\_intercept & & Intercept of the relation \\
age\_dm\_scatter & & Scatter of the relation \\
\multicolumn{3}{c}{Distance modulus - logarithm of mass relation:} \\ 
dm\_mass\_slope & & Slope of the relation \\
dm\_mass\_intercept & & Intercept of the relation \\
dm\_mass\_scatter & & Scatter of the relation \\
\bottomrule
\end{tabular}
\end{center}
\caption{Measures in the output table. For items denoted by a cross ($\dagger$) multiple columns are provided; see section \ref{sec:outvalues} for details. More than one USPDF (i.e. more than one line) can be reported for each stage of each star.}\label{tbl:output}
\end{table}

\subsection{Role of age and extinction cuts}\label{sec:cuts}
We chose to impose hard cuts on log(age) $\tau \leq 10.13$ and extinction $A_K \geq 0$. This is not an obvious choice, so we justify it here.

\begin{figure}
  \myimagesmall{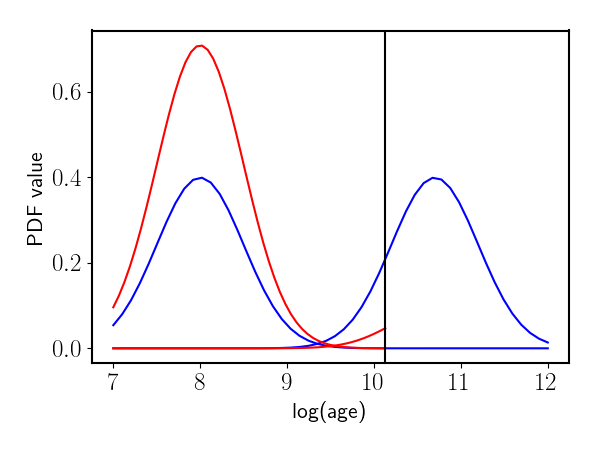}
  \caption{Sketch of PDF in log(age) for a star with two possible ages without (blue lines) and with (red lines) hard upper limit on log(age). See text for details.}\label{fig:cut}
\end{figure}

We first consider a star for which two equally good solutions are possible: one with $\tau = 8$ and one with $\tau = 10.7$, where all other parameters are equal (see \autoref{fig:cut}). If we do not use the hard cut on ages, both solutions are reported with good quality flags and equal weights $W_{1,2} = 0.5$ (blue lines in \autoref{fig:cut}). 
The unphysical age value for the second solution might be used as a sign of a problem either with data or with models. It is therefore likely that this solution or even both solutions for this star will be dropped from further study. 
If we, on the other hand, use the cut in log(age) $\tau \leq 10.13$, we will keep both solutions, but their weights will change. Only the tail of the USPDF for the second solution will be retained. Because the sum of USPDF weight still has to be unity, the weight of the first solution will increase. We get for this example $W_1 = 0.89$ and $W_2 = 0.11$ (red lines in \autoref{fig:cut}). As a result, we get a ``realistic'' first solution with $\tau = 8$ and retain a part of the second solution. For the second solution, and, in general, for all cases when part of the USPDF is cut away, the mean of the USPDF is a poor measure of log(age) and it is biased towards lower values. But in such cases the USPDF is fitted with either a MTED, a truncated Gaussian or a truncated \stud. So instead of a solution with high weight and correct, but unphysical, mean log(age), we get a solution with lower weight and biased value of the mean with a proper fitting function.

We now consider the case when only one solution is available for a given star, which is $\tau = 10.7$. 
We can use such values of $\tau$ as an indication of a problem in spectroscopic parameters, photometry, or in isochrones. 
If the cut in log(age) $\tau \leq 10.13$ is applied, there are several possibilities. In some cases the PDF in log(age) follows an exponential distribution, which means that the age is poorly constrained for a given star and extending the range of possible ages does not improve this.

An other possibility is that models that have $\tau \leq 10.13$ represent observables and thus have $p_{best}$ close to unity; see \autoref{sec:p_best_p_sed}. This means that we still have a reliable log(age) PDF for $\tau \leq 10.13$, but the mean, mode, and median values might be biased. Without the age cut, the mean, mode, and median log(age) values will be above $\tau = 10.13$, and such solution will likely be excluded from further analysis, despite the fact that a fraction of it is reliable.

In yet another case the value for a best-fitting model probability $p_{best}$ will be small, which will indicate potential problems in either the data or with the models. Such cases will be flagged as unreliable with and without age cut.

The same arguments as discussed above are applicable to the cut in extinction. Moreover, a cut in extinction has minor influence on the result, as extinction values are typically very small, i.e. about 10 times smaller than the derived uncertainty in the distance modulus. We verified that negative extinctions typically arise for faint stars, for which photometric uncertainties are large. In the vast majority of cases for which the derived value of extinction is negative, the value is still consistent with zero within uncertainties.

\section{Tests: Comparison with other measurements}\label{sec:test}
We tested our UniDAM tool in two ways.
We first applied it to mock surveys (see Section \ref{sec:data_mock}). By doing that we checked the accuracy and performance of the tool
and explored the effect of random perturbations added to input values.
Then we proceeded with comparing our parameter measurements for real stars with those obtained by other groups and presented in the literature. 
The aim of these exercises was to check the quality of our estimates compared to results obtained by the 
consortia of the different surveys and the sensitivity of our results to priors.

\subsection{Mock survey}
We ran our UniDAM tool on all four mock surveys, described in Section \ref{sec:data_mock}. Knowing the input values allowed us to evaluate which of the reported measures, that is mean, median, or mode, is the best proxy. In agreement with \cite{2005A&A...436..127J} we find that the mode is less biased than the mean or median, but produces slightly more outliers.

Mean, mode, and median values show similar qualitative patterns, so we used only mean output values of the highest weight USPDF for comparison.
We compared derived mean values $X$ with input values $X_0$. We considered several measures of interest as follows: \begin{itemize}
 \item Median fractional uncertainty of derived value $\median{\frac{\sigma_X}{X}}$ ; this is an internal precision measure.
 \item Median relative deviation (median bias) of derived value $\median{\frac{\Delta X}{X}} = \median{\frac{X_0-X}{X}}$; this shows whether the values that we calculate are systematically offset with respect to the input.
 \item Median absolute relative deviation of derived value $\MAD = \median{\left|\frac{\Delta X}{X} - \median{\frac{\Delta X}{X}}\right|}$; this shows how scattered our derived values are with respect to input. Median absolute deviation is a much better estimate of scatter than standard deviation in the presence of outliers \citep{Leys2013}.
 \item Outlier fraction rate $O$; this is the fraction of stars for which the input value $X_0$ lies outside the three-sigma confidence interval. We use two values: $O_{best}$ is calculated using only highest weight USPDF, whereas $O_{all}$ is calculated using all USPDFs (i.e. $X_0$ lies outside the three-sigma confidence intervals of all reported USPDFs). 
\end{itemize}

Measures for all four mock surveys are listed in \autoref{tbl:mock}. 
For a normally distributed random variable, median absolute deviation $\sigma_{MAD}$ relates to standard deviation $\sigma$ as $\sigma \approx 1.4826\,\sigma_{MAD}$, therefore we expected median absolute relative deviation to be approximately equal to two-thirds of the median fractional uncertainty. As can be seen from the second and fourth columns of \autoref{tbl:mock}, this is the case for our data, which means that the distribution of the offsets is close to normal.

Outliers are expected for two reasons. First, a model with the "correct" (i.e. input) values might not belong to the highest weight USPDF. This is revealed by $O_{best}$. We detected that in about one to seven percent of the cases input parameters are better recovered with USPDF that has second (or even the third) priority. This happens primarily in the upper part of the giant branch, where isochrone overlap is highest. This is inherent to the method; we seek a model that most likely represents the data. If the mock star was taken to be in some short phase of its evolution (thus having low model weight $w_j$), chances are high that we assign a highest weight USPDF not to this phase but to a much longer phase, which has similar observables. Second, because the three-sigma range includes by definition $99.7\%$ of the data, we expect at least a fraction of $0.003$ of stars for which no USPDF recovers the ``correct'' values within three-sigma confidence intervals (i.e. $O_{all} \gtrsim 0.003$) due to our random perturbations added to input values. In fact, this fraction is slightly higher, of the order of $0.01-0.02$. We checked that this is caused by a combination of both the perturbations of input values and models selected for the mock sample that is in a very short phase of evolution. In the latter case USPDFs might be pulled away from the ``correct'' solution by nearby models with higher weights. Another possible case in which there might be no ``correct'' solution found is when the mock star is located on the edge of the parameter space covered by models.

In cases where perturbations were added, the values of the bias, median absolute relative deviation, and outlier fractions increase. This increase is most prominent in the outlier fractions; this is caused by the fact that sometimes even a small perturbation of input parameters might change the priorities of USPDFs, thus changing the parameters of the highest weight USPDF by a large value. 

\begin{table}
\begin{center}
\begin{tabular}{lrrrp{0.9cm}p{0.9cm}} \toprule
Parameter & $\median{\frac{\sigma_X}{X}}$ & $\median{\frac{\Delta X}{X}}$ & $\MAD$ & $O_{best}$ & $O_{all}$ \\ \midrule
\multicolumn{6}{c}{No perturbation} \\
mass & 0.17 & -0.02 & 0.10 & 0.06 & 0.01 \\
age & 0.03 & 0.01 & 0.02 & 0.06 & 0.00 \\
distance & 0.13 & -0.02 & 0.08 & 0.06 & 0.00 \\
\multicolumn{6}{c}{Photometry perturbation} \\
mass & 0.17 & -0.00 & 0.10 & 0.04 & 0.01 \\
age & 0.03 & 0.00 & 0.02 & 0.04 & 0.00 \\
distance & 0.13 & 0.01 & 0.08 & 0.03 & 0.00 \\
\multicolumn{6}{c}{Spectral parameters perturbation} \\
mass & 0.17 & -0.00 & 0.12 & 0.06 & 0.02 \\
age & 0.03 & 0.00 & 0.02 & 0.06 & 0.01 \\
distance & 0.13 & 0.00 & 0.11 & 0.05 & 0.01 \\
\multicolumn{6}{c}{Photometry and spectral parameters perturbation} \\
mass & 0.17 & -0.01 & 0.12 & 0.06 & 0.02 \\
age & 0.03 & 0.01 & 0.02 & 0.06 & 0.01 \\
distance & 0.13 & 0.00 & 0.11 & 0.06 & 0.01 \\
\bottomrule
\end{tabular}
\caption{Results of mock-survey comparison. $\median{t}$ is the median value of $t$; $\Delta X = X_0 - X$ is the difference between input mock survey value $X_0$ and our estimate $X$;  and $\sigma_X$ is the uncertainty of our estimate. 
$O_{best}$ and $O_{all}$ are outlier fractions calculated using the highest weight USPDF and all USPDFs, respectively. See text for details.}\label{tbl:mock}
\end{center}
\end{table}

We also tested if distance modulus or parallax provide a better estimate of distance than distance value itself. We find that this is not the case, and all three estimates give very similar precision, with distance itself showing a slightly smaller fraction of outliers. This contradicts the statement of \cite{2014MNRAS.437..351B} that ``the most reliable distance indicator is the expectation of parallax''. This might be because \cite{2014MNRAS.437..351B} compared their derived values with parallaxes from \textit{HIPPARCOS}, and comparing parallaxes with parallaxes is likely less biased.
We nevertheless provide all three estimates for each star in the output catalogue.

\subsection{Literature}\label{sec:compare}
\begin{figure*}
 \myimageH{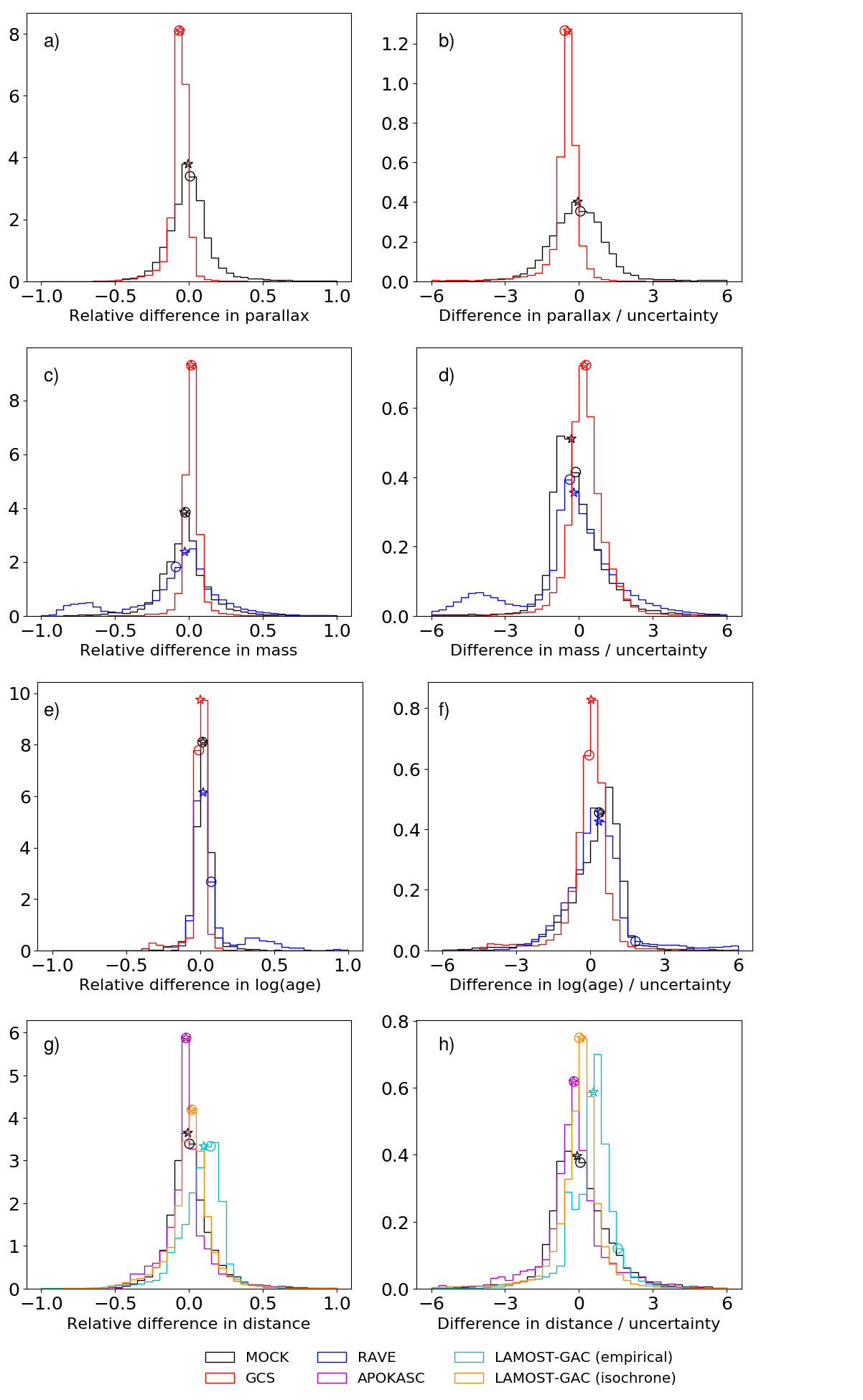}
 \caption{Distributions of relative difference $\frac{X_0-X}{X}$ (left) and difference in units of uncertainty $\frac{X_0-X}{\sigma_X}$ (right) for values provided in the literature $X_0$ (see legend) and our results $X$. Circles indicate mean values, star symbols indicate median values. For RAVE survey values from \cite{2014MNRAS.437..351B} were used for comparison.} \label{fig:Compare_input}
\end{figure*}

We compared our results with values available in the literature. Results of this comparison are shown in \autoref{fig:Compare_input} and in \autoref{tbl:Compare_input}, and are discussed here. 
The aim of this comparison is to show that our results are consistent with previous studies. In most cases these previous studies were based on similar data and methods. So the differences that appear
are primarily due to different models used and differences in details of the method implementation. The exceptions are GCS parallaxes that are coming from \textit{HIPPARCOS} and APOKASC distances, derived with asteroseismic values of $\log g$, which are more precise than spectroscopic values. In both cases our results are consistent with published data.

We verified that our extinction estimates are consistent with those provided in surveys. In fact, the differences between our extinction estimates and those in surveys are comparable to differences between extinctions derived with different methods for the same survey, for example, in the LAMOST-GAC data \citep{2017arXiv170105409X}. We do not provide a detailed analysis here, as the derivation of precise extinctions is beyond the scope of this work.

\begin{table}
\centering
\begin{tabular}{ccd{1.2}d{1.2}d{1.2}c}
\toprule
Survey     & Value        & \MyHead{1cm}{$\median{\frac{\sigma_X}{X}}$} & \MyHead{0.9cm}{$\median{\frac{\Delta X}{X}}$} & \MyHead{0.9cm}{$\MAD$} & $O_{best}$ \\ \midrule
GCS        &$\pi$              & 0.13  & -0.065  & 0.03   & 0.03 \\
           &$\mass$            & 0.064 & 0.024   & 0.024  & 0.02 \\
           &$\tau$             & 0.02  & -0.000  & 0.005  & 0.05 \\ \midrule
APOKASC    &$d$                & 0.11  & -0.017  & 0.05   & 0.15 \\ \midrule
RAVE (1)   & $\mu_{d, \textrm{Z}}$ & 0.044 & -0.011   & 0.050  & 0.20 \\ \midrule
RAVE (2)   &$\mu_{d, \textrm{B}}$  & 0.044 & -0.03   & 0.055  & 0.25 \\
           &$\mass$            & 0.15  & -0.016  & 0.12   & 0.17 \\
           &$\tau$             & 0.027 & 0.009   & 0.018  & 0.13 \\ \midrule
LAMOST-GAC &$d_{\textrm{emp}}$ & 0.20  & 0.1    & 0.08    & 0.03 \\
(main sample)&$d_{\textrm{iso}}$ & 0.20  & 0.02   & 0.063   & 0.05 \\ \bottomrule
\end{tabular}
\caption{Differences between values provided in the literature and our results. $\median{t}$ is the median value of $t$. $\Delta X = X_0 - X$ is the difference between input value $X_0$ and our estimate $X$, $\sigma_X$ is the uncertainty of our estimate.
$\MAD$ is the median absolute deviation of $\frac{\Delta X}{X}$, $O_{best}$ is the outlier fraction calculated using highest-weight USPDF. For RAVE comparison is done for two publications: (1) -- \cite{2010A&A...522A..54Z}, (2) -- \cite{2014MNRAS.437..351B}. See text for details.} \label{tbl:Compare_input}
\end{table}

\subsubsection{GCS parallaxes, masses, and ages}
The GCS \citep[][and references therein]{2011AA...530A.138C} mainly covers nearby main-sequence stars. The big advantage is that for most of them parallaxes were measured by \textit{HIPPARCOS}. 
The three top rows of \autoref{tbl:Compare_input} and \autoref{fig:Compare_input} show differences between parallaxes, masses, and log(age)s from GCS and our estimate.
We detected a small bias in parallaxes and masses but a negligible bias in log(age)s. Median absolute deviations are three times lower than fractional uncertainties, which means that our method is consistent with GCS results.

\subsubsection{Distances of APOKASC red giants}\label{sec:compare_apokasc}
\cite{2014MNRAS.445.2758R} have determined distances for about 2000 red giant stars from the APOKASC sample.  Our distances are less precise than those of \cite{2014MNRAS.445.2758R}  because we do not include asteroseismic data. This test therefore helps us estimate the quality of distance estimations we make for the whole APOGEE sample, as stellar parameters in APOGEE DR13 were calibrated with the use of asteroseismic data from APOKASC. We predicted slightly larger distances ($-0.017$ relative offset), but both bias and scatter are well below the mean fractional uncertainty ($0.11$) of our derived distances. The origin of the bias is likely the difference in how the distance value is calculated when distance PDF is multimodal. This is supported by the fact that for stars with unimodal PDFs we got a relative distance bias of less than $0.0025$.

\subsubsection{RAVE stars}\label{sec:compare_rave}
We ran the UniDAM tool on RAVE DR4 data \citep{2013AJ....146..134K} and compared these findings with the results of \citet{2010A&A...522A..54Z} for distance moduli and \citet{2014MNRAS.437..351B} for distance moduli, log(age)s, and masses. We used DR4 data here, as these were used by \citet{2010A&A...522A..54Z} and \citet{2014MNRAS.437..351B}. The relative difference between our distance estimates and $\mu_{d, \textrm{Z}}$ by \citet{2010A&A...522A..54Z} 
is around $-0.01$. As compared to the \citet{2014MNRAS.437..351B} results ($\mu_{d, \textrm{B}}$), our distance moduli have a relative difference of $-0.03$. The median absolute deviations are large in both cases and are comparable to or larger than the mean relative uncertainties of our values.

The reason for a larger difference for $\mu_{d, \textrm{B}}$ is that \citet{2014MNRAS.437..351B} use strong priors on distances, metallicities, and ages 
coming from a model of the Galaxy. These priors are decreasing functions of distance from the Galactic centre and from the Galactic plane. Therefore they decrease with distance from the Sun for the majority of directions probed by RAVE. The prior that decreases with distance results in smaller estimates for stellar distances and thus slightly smaller masses and larger ages as compared to our results. Difference in log(age) are further enhanced due to age priors used by \citet{2014MNRAS.437..351B}.

As can be seen in panels \textit{c} and \textit{e} of \autoref{fig:Compare_input}, distributions of differences between our results for log(age)s and masses and \citet{2014MNRAS.437..351B} results are bimodal, with a secondary peak at approximately $-0.75$ in relative mass difference and $0.2$ in relative log(age) difference. The same can be seen in panel \textit{d}. In panel \textit{f} the second peak is out of the plotted range.
This second peak contains about 12\% of stars and is caused by a difference in the evolutionary stages accepted in \citet{2014MNRAS.437..351B} and by our UniDAM tool. A similar pattern but with a much smaller secondary peak can be seen with data from the mock survey (black histogram in panels c and d of \autoref{fig:Compare_input}).

\begin{figure}
 \myimagesmall{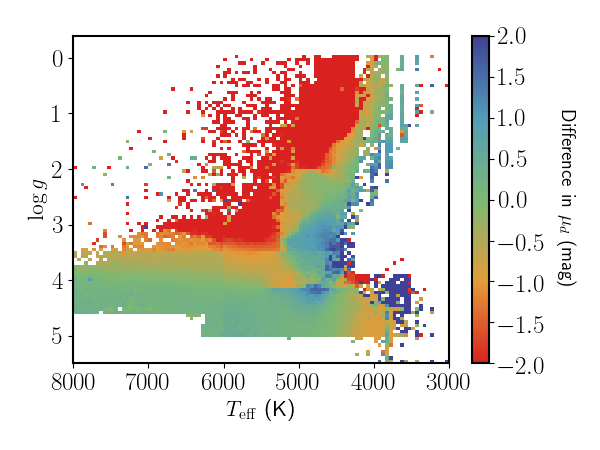}
 \myimagesmall{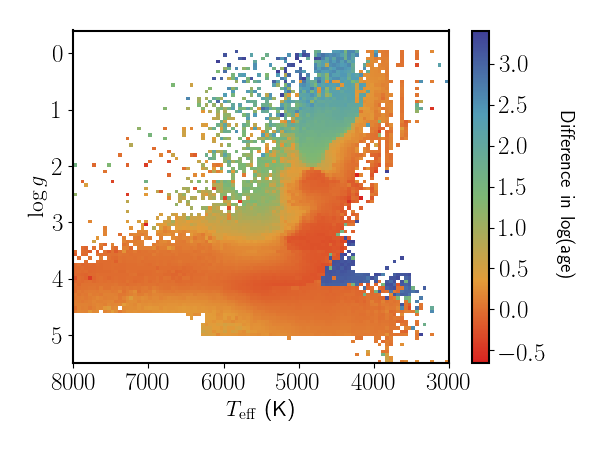}
 \caption{Hertzsprung-Russell diagrams of RAVE data showing colour differences between our results and RAVE results \citep{2014MNRAS.437..351B} for distance moduli (top panel) and log(age)s (bottom panel).} \label{fig:RAVE}
\end{figure}

We show in \autoref{fig:RAVE} the distributions of the median difference between our results and the \citet{2014MNRAS.437..351B} results for distance modulus and log(age) on the Hertzsprung-Russell diagram.
We chose RAVE because it contains estimates for both distance modulus and log(age) and because it covers both main-sequence and giant stars. 
There is clearly a good agreement in both distance modulus and log(age) for the main-sequence stars and large fraction of giant branch, including the red clump. 
A disagreement for pre-main-sequence stars and large and hot (thus most massive) giants is primarily due to difference in the models and priors used. Similar plots can be produced for other datasets, revealing similar patterns. 

\subsubsection{LAMOST-GAC distances}
We compared our results with two distance estimates provided in \cite{2015RAA....15.1095L}. Our values are systematically smaller by a fraction of $0.1$ as compared to their ``empirical'' estimates based on the MILES library. We have much better agreement with estimates based on isochrones from Dartmouth Stellar Evolution Database ($0.02$ fractional difference). \cite{2015RAA....15.1095L} do not provide uncertainties for their distance estimates.
Relative uncertainties of our distance estimates for LAMOST-GAC are higher than estimates build on data from other surveys due to the higher uncertainties in spectral parameters, which lead to a fractional uncertainties on our distances of $0.2$.

\subsection{Effect of the volume correction}\label{sec:priors}
We ran tests to see how much the use of the volume correction (see \autoref{eq:model_weight}) affects our results.
Volume correction can be seen as a distant prior that ensures constant number density.
In general, if the distance prior is a decreasing (increasing) function, the resulting distance is smaller (larger) than in the case of a flat prior. The size of this effect depends on the relative variation of the prior function within the uncertainty range of the parameter. 
We chose two datasets for the test: GCS and APOKASC giants \citep{2014MNRAS.445.2758R}.
The GCS dataset contains primarily main-sequence stars with distances derived from \textit{HIPPARCOS} parallaxes. These parallaxes are in most cases more precise than our distance measurements. Distances of APOKASC giants were derived using asteroseismic data, and therefore should also be more precise than our measurements. We ran our UniDAM tool with and without volume correction. 
We selected only USPDFs with highest weight for analysis in each case. We then explored how parallaxes, distances, masses, and log(age)s were affected by volume correction.
For multimodal cases it is important that the use of the volume correction  might change the relative weights of USPDFs, so that priorities might also change.
The result of our experiment is that in $7\%$ cases for APOGEE the assigned evolutionary stage changed when we applied the volume correction. This did not happen for GCS as PDFs are unimodal in most cases for main-sequence stars in that survey.
By removing volume correction we decreased distance estimations in both datasets by a fraction of $0.032$ if the assigned evolutionary stage did not change. This is well below the median relative distance uncertainties that we have ($\approx 0.13$). The mass estimates are correlated with distance, and decreased by a fraction of $0.03$, again, this is well below relative uncertainties in mass that we find ($\approx 0.15$). The logarithm of age estimates, which are anti-correlated with distance, increased, but only by a fraction of $0.005$  (log(age) fractional uncertainties $\approx 0.03$). So the conclusion here is that the volume correction has a measurable and well understood 
effect on measured parameters, but this effect is smaller than our typical parameter uncertainties.
This effect is systematic and has to be taken into account when comparing with results obtained with distance priors; see for example \autoref{sec:compare_rave}. However, we expect the contribution from the (unknown) systematic uncertainties of spectroscopic measurements to be at least as high as the influence of the volume correction.

We also compare how the volume correction affects the agreement between our measurements and data from the literature, which is described above in Section \ref{sec:compare}. The effect of volume correction is summarised in the \autoref{tab:prior}.
For the GCS sample there is a clear advantage of using the volume correction. 
Without volume correction our parallax estimates are lower than those in GCS by a fraction $0.091$.
If we use the volume correction, our parallax estimates increase on average, which improves the agreement (fractional difference of $0.059$; see the first row of \autoref{tab:prior}). The same applies for log(age) and mass estimates.
As for the APOKASC sample, there seems to be an opposite result, as we seem to overestimate the distance compared to \cite{2014MNRAS.445.2758R}. This is likely caused by the fact that distances provided in \cite{2014MNRAS.445.2758R} are in fact modes of probability density function. If we use modes instead of means for our USPDFs for distances, than we get a relative difference of less than $10^{-3}$ if the volume correction is included and a relative difference around $0.014$ if there is no volume correction used.

The effect of the volume correction increases gradually with increasing distances, where our distance uncertainty is larger. This is caused by an increase in the relative variation of the value of the volume correction within the distance uncertainty. The effect of volume correction is approximately proportional to a square of the uncertainty in distance modulus. 
If the assigned evolutionary stage changes, the estimates of distance, mass, and log(age) can change by a large amount, sometimes by more than $50\%$.

\begin{table}
\centering
\begin{tabular}{ccd{1.3}d{1.3}}
\toprule
Survey & Value & \MyHead{1.7cm}{ $\median{\frac{\Delta X}{X}}$ without volume correction} & \MyHead{1.7cm}{ $\median{\frac{\Delta X}{X}}$ with volume correction}\\ \midrule
GCS    & $\pi$  & -0.091 & -0.059 \\ 
       & $M$   & 0.027 & 0.015  \\
       & $\tau$ & -0.000 &  0.000  \\ \midrule
APOKASC & $d$  & -0.001 & -0.021  \\ \bottomrule
\end{tabular}
\caption{Value of $\median{\frac{\Delta X}{X}}$ for measures without and with the volume correction applied.}\label{tab:prior}
\end{table}

\section{Stellar parameters catalogue}
\begin{figure*}
 \myimage{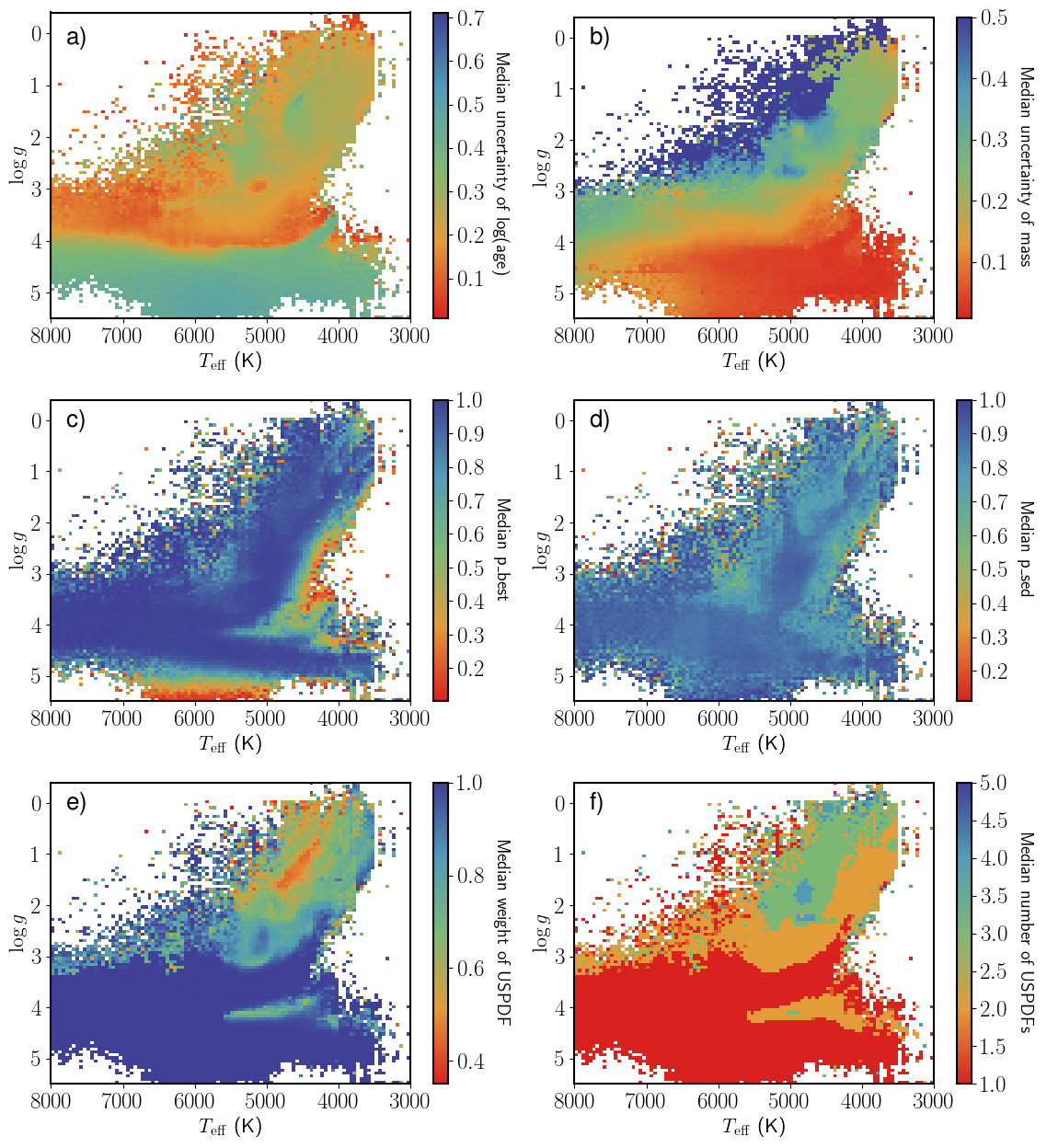}
 \caption{Median values across the catalogue: a) median uncertainty in log(age), b) median uncertainty in mass, c) $p_{best}$, d) $p_{sed}$,  e) median weight of the highest weight USPDF, and f) median number of USPDFs with weight $V>0.03$ per star. } \label{fig:Catalog}
\end{figure*}

We provide a catalogue of stellar distances, masses, and log(age)s determined with the UniDAM tool described in this manuscript. 
Our catalogue contains over 3.8 million rows (one row for each USPDF) for over 2.5 million stars. 
We summarise some properties of this catalogue in \autoref{fig:Catalog}. This figure shows medians
of different quantities in each bin on the Hertzsprung-Russell (HR) diagram. 
Data from all input spectroscopic surveys have been used to produce this figure. Quantification of differences between spectroscopic data from different surveys and effects of incompleteness and selection are beyond the scope of this paper and will be addressed in future work. Here, we are interested in a qualitative description of how the quality of our estimates vary in different parts of the HR diagram.
Panels \textit{a} and \textit{b} of \autoref{fig:Catalog} show uncertainties in measured log(age)s and masses. Log-ages are best constrained 
in the upper part of the main sequence, where parameters change fast as a star leaves the main sequence. 
On the contrary, masses are much better constrained on the main sequence.

Panels \textit{c} and \textit{d} show median values of $p_{best}$ (the probability for a best-fitting model) and $p_{sed}$ (a measure of how good we reproduce SED with our model). Patterns on both panels are similar with worse results close to the edges of the region covered by the PARSEC isochrones and additionally for $p_{sed}$ between the main sequence and giant branch.

Panels \textit{e} and \textit{f} show median values for the weights $V_0$ of the highest weight USPDF and total number of USPDFs with $V_i > 0.03$. The patterns are nearly inverse: on the main sequence we typically have only one USPDF with weight equals to unity, whereas on the giant branch the number of USPDFs increases and the weights of the highest weight USPDF decreases. It is important that for the giant branch we typically have two or more USPDFs, therefore using just the one with the highest weight is insufficient; the best solution is to use all USPDFs with their relative weight taken into account.

\subsection{Quality flags}\label{sec:quality}
The output catalogue contains a \texttt{quality} column, which indicates how reliable data contained in each row are. Values have been assigned as follows:
\begin{description}
\item[1] - single PDF
\item[A] - highest-weight USPDF has power of 0.9 or more
\item[B] - 1st and 2nd priority USPDFs together have power of 0.9 or more
\item[C] - 1st, 2nd, and 3rd priority USPDFs together have power of 0.9 or more
\item[D] - 1st, 2nd, and 3rd priority USPDFs together have power of less than 0.9
\item[L] - low power USPDF (between 0.03 and 0.1)
\item[E] - USPDF has $p_{sed} < 0.1$ (possibly bad photometry)
\item[X] - highest weight USPDF has $p_{best} < 0.1$ (likely off the model grid)
\item[N] - USPDF has less than 10 models (unreliable result)
\end{description}
Although \texttt{the quality} value provides some information on the quality of the parameter estimation, it is not recommended to select stars based on that value alone (apart from removing unreliable results with values \textbf{E, N,} or \textbf{X}), because the quality value varies heavily over the HR diagram: for main-sequence stars the quality is in most cases \textbf{1} or \textbf{A}, whereas for giants quality  \textbf{B, C,} or even \textbf{D} are much more common. This is illustrated by the distribution of the number of USPDFs in panel f of \autoref{fig:Catalog}. There are $2\%$ cases where a highest weight USPDF has quality \textbf{E}, $4.2\%$ cases with quality \textbf{X,} and less than $0.01\%$ cases with quality \textbf{N}.

\section{Discussion and conclusions}\label{sec:discussion}
We provide a catalogue of distances, log(age)s, and masses for over 2.5 million stars. This number will increase as new data is made available, for example new data releases for surveys already included, or data from new surveys. Gaia data will be of high value and can be used as an independent test of our distances or as a parallax prior. In the latter case it should improve our extinction, mass, and log(age) estimates considerably.

In the current version of our UniDAM tool we use infrared magnitudes, \threeparams\ as inputs to derive distances, log(age)s, and masses of stars. The tool was also successfully used to derive temperatures for a APOKASC sample, with inputs being surface gravities, and masses derived from seismic information and spectroscopic metallicities (Tayar et al. 2017, accepted).

An advantage of our approach is that we represent multi-peaked PDFs for parameters with a sum of unimodal distributions. Additionally we provide parameters of fits representing each distribution and the correlations between distance modulus, log(age), and mass. Therefore our catalogue contains not only mean values and uncertainties, but detailed information on PDFs. This allows us to apply more sophisticated analysis to the dataset to reveal both global and local structures in the Galaxy.

The next step will be to add proper motion data, thus obtaining all six dimensions of stellar positions and velocities. Combination of positions and velocities with ages, metallicities, and (where available) chemical abundances will open up new possibilities to study Galactic structure. Furthermore, it is important to get a correct estimate of the selection function, as this might affect results not only quantitatively, but also qualitatively, as was shown by \citet{2012ApJ...751..131B}. We intend to produce a selection function for our catalogue and then proceed to study Galactic structure on large and small scales.

\section*{Acknowledgements}
Authors thank the anonymous referee for a detailed report with many useful suggestions. It helped us to improve the manuscript substantially.

The research leading to the presented results has received funding from the European Research Council under the European Community's Seventh Framework Programme (FP7/2007- 2013)/ERC grant agreement (No 338251, StellarAges). 

This publication makes use of data products from the Two Micron All Sky Survey, which is a joint project of the University of Massachusetts and the Infrared Processing and Analysis Center/California Institute of Technology, funded by the National Aeronautics and Space Administration and the National Science Foundation. 

Guoshoujing Telescope (the Large Sky Area Multi-Object Fiber Spectroscopic Telescope LAMOST) is a National Major Scientific Project built by the Chinese Academy of Sciences. Funding for the project has been provided by the National Development and Reform Commission. LAMOST is operated and managed by the National Astronomical Observatories, Chinese Academy of Sciences. 

This publication makes use of data products from the Wide-field Infrared Survey Explorer, which is a joint project of the University of California, Los Angeles, and the Jet Propulsion Laboratory/California Institute of Technology, and NEOWISE, which is a project of the Jet Propulsion Laboratory/California Institute of Technology. WISE and NEOWISE are funded by the National Aeronautics and Space Administration.

Funding for RAVE has been provided by the Australian Astronomical Observatory; the Leibniz-Institut fuer Astrophysik Potsdam (AIP); the Australian National University; the Australian Research Council; the French National Research Agency; the German Research Foundation (SPP 1177 and SFB 881); the European Research Council (ERC-StG 240271 Galactica); the Istituto Nazionale di Astrofisica at Padova; The Johns Hopkins University; the National Science Foundation of the USA (AST-0908326); the W. M. Keck foundation; the Macquarie University; the Netherlands Research School for Astronomy; the Natural Sciences and Engineering Research Council of Canada; the Slovenian Research Agency; the Swiss National Science Foundation; the Science and Technology Facilities Council of the UK; Opticon; Strasbourg Observatory; and the Universities of Groningen, Heidelberg and Sydney. The RAVE website is at \url{https://www.rave-survey.org}. 

Funding for SDSS-III has been provided by the Alfred P. Sloan Foundation, the Participating Institutions, the National Science Foundation, and the U.S. Department of Energy Office of Science. The SDSS-III website is http://www.sdss3.org/. SDSS-III is managed by the Astrophysical Research Consortium for the Participating Institutions of the SDSS-III Collaboration including the University of Arizona, the Brazilian Participation Group, Brookhaven National Laboratory, University of Cambridge, Carnegie Mellon University, University of Florida, the French Participation Group, the German Participation Group, Harvard University, the Instituto de Astrofisica de Canarias, the Michigan State/Notre Dame/JINA Participation Group, Johns Hopkins University, Lawrence Berkeley National Laboratory, Max Planck Institute for Astrophysics, Max Planck Institute for Extraterrestrial Physics, New Mexico State University, New York University, Ohio State University, Pennsylvania State University, University of Portsmouth, Princeton University, the Spanish Participation Group, University of Tokyo, University of Utah, Vanderbilt University, University of Virginia, University of Washington, and Yale University.

This research has made use of the SIMBAD database, operated at CDS, Strasbourg, France

\bibliographystyle{aa.bst}
\bibliography{sage_gap.bib}

\begin{thebibliography}{70}
\expandafter\ifx\csname natexlab\endcsname\relax\def\natexlab#1{#1}\fi

\bibitem[{{Ahn} {et~al.}(2014){Ahn}, {Alexandroff}, {Allende Prieto}, {Anders},
  {Anderson}, {Anderton}, {Andrews}, {Aubourg}, {Bailey}, {Bastien}, \&
  et~al.}]{2014ApJS..211...17A}
{Ahn}, C.~P., {Alexandroff}, R., {Allende Prieto}, C., {et~al.} 2014, \apjs,
  211, 17

\bibitem[{{Alam} {et~al.}(2015){Alam}, {Albareti}, {Allende Prieto}, {Anders},
  {Anderson}, {Anderton}, {Andrews}, {Armengaud}, {Aubourg}, {Bailey}, \&
  et~al.}]{2015ApJS..219...12A}
{Alam}, S., {Albareti}, F.~D., {Allende Prieto}, C., {et~al.} 2015, \apjs, 219,
  12

\bibitem[{{Allende Prieto} {et~al.}(2008){Allende Prieto}, {Sivarani}, {Beers},
  {Lee}, {Koesterke}, {Shetrone}, {Sneden}, {Lambert}, {Wilhelm}, {Rockosi},
  {Lai}, {Yanny}, {Ivans}, {Johnson}, {Aoki}, {Bailer-Jones}, \& {Re
  Fiorentin}}]{2008AJ....136.2070A}
{Allende Prieto}, C., {Sivarani}, T., {Beers}, T.~C., {et~al.} 2008, \aj, 136,
  2070

\bibitem[{{Baglin} {et~al.}(2006){Baglin}, {Auvergne}, {Barge}, {Deleuil},
  {Catala}, {Michel}, {Weiss}, \& {COROT Team}}]{2006ESASP1306...33B}
{Baglin}, A., {Auvergne}, M., {Barge}, P., {et~al.} 2006, in ESA Special
  Publication, Vol. 1306, The CoRoT Mission Pre-Launch Status - Stellar
  Seismology and Planet Finding, ed. M.~{Fridlund}, A.~{Baglin}, J.~{Lochard},
  \& L.~{Conroy}, 33

\bibitem[{{Binney} {et~al.}(2014){Binney}, {Burnett}, {Kordopatis}, {McMillan},
  {Sharma}, {Zwitter}, {Bienaym{\'e}}, {Bland-Hawthorn}, {Steinmetz},
  {Gilmore}, {Williams}, {Navarro}, {Grebel}, {Helmi}, {Parker}, {Reid},
  {Seabroke}, {Watson}, \& {Wyse}}]{2014MNRAS.437..351B}
{Binney}, J., {Burnett}, B., {Kordopatis}, G., {et~al.} 2014, \mnras, 437, 351

\bibitem[{{Borucki} {et~al.}(2010){Borucki}, {Koch}, {Basri}, {Batalha},
  {Brown}, {Caldwell}, {Caldwell}, {Christensen-Dalsgaard}, {Cochran},
  {DeVore}, {Dunham}, {Dupree}, {Gautier}, {Geary}, {Gilliland}, {Gould},
  {Howell}, {Jenkins}, {Kondo}, {Latham}, {Marcy}, {Meibom}, {Kjeldsen},
  {Lissauer}, {Monet}, {Morrison}, {Sasselov}, {Tarter}, {Boss}, {Brownlee},
  {Owen}, {Buzasi}, {Charbonneau}, {Doyle}, {Fortney}, {Ford}, {Holman},
  {Seager}, {Steffen}, {Welsh}, {Rowe}, {Anderson}, {Buchhave}, {Ciardi},
  {Walkowicz}, {Sherry}, {Horch}, {Isaacson}, {Everett}, {Fischer}, {Torres},
  {Johnson}, {Endl}, {MacQueen}, {Bryson}, {Dotson}, {Haas}, {Kolodziejczak},
  {Van Cleve}, {Chandrasekaran}, {Twicken}, {Quintana}, {Clarke}, {Allen},
  {Li}, {Wu}, {Tenenbaum}, {Verner}, {Bruhweiler}, {Barnes}, \&
  {Prsa}}]{2010Sci...327..977B}
{Borucki}, W.~J., {Koch}, D., {Basri}, G., {et~al.} 2010, Science, 327, 977

\bibitem[{{Bovy} {et~al.}(2012){Bovy}, {Rix}, \& {Hogg}}]{2012ApJ...751..131B}
{Bovy}, J., {Rix}, H.-W., \& {Hogg}, D.~W. 2012, \apj, 751, 131

\bibitem[{{Breddels} {et~al.}(2010){Breddels}, {Smith}, {Helmi},
  {Bienaym{\'e}}, {Binney}, {Bland-Hawthorn}, {Boeche}, {Burnett}, {Campbell},
  {Freeman}, {Gibson}, {Gilmore}, {Grebel}, {Munari}, {Navarro}, {Parker},
  {Seabroke}, {Siebert}, {Siviero}, {Steinmetz}, {Watson}, {Williams}, {Wyse},
  \& {Zwitter}}]{2010A&A...511A..90B}
{Breddels}, M.~A., {Smith}, M.~C., {Helmi}, A., {et~al.} 2010, \aap, 511, A90

\bibitem[{{Bressan} {et~al.}(2012){Bressan}, {Marigo}, {Girardi}, {Salasnich},
  {Dal Cero}, {Rubele}, \& {Nanni}}]{PARSEC}
{Bressan}, A., {Marigo}, P., {Girardi}, L., {et~al.} 2012, \mnras, 427, 127

\bibitem[{{Burnett} {et~al.}(2011){Burnett}, {Binney}, {Sharma}, {Williams},
  {Zwitter}, {Bienaym{\'e}}, {Bland-Hawthorn}, {Freeman}, {Fulbright},
  {Gibson}, {Gilmore}, {Grebel}, {Helmi}, {Munari}, {Navarro}, {Parker},
  {Seabroke}, {Siebert}, {Siviero}, {Steinmetz}, {Watson}, \&
  {Wyse}}]{2011A&A...532A.113B}
{Burnett}, B., {Binney}, J., {Sharma}, S., {et~al.} 2011, \aap, 532, A113

\bibitem[{{Carlin} {et~al.}(2015){Carlin}, {Liu}, {Newberg}, {Beers}, {Chen},
  {Deng}, {Guhathakurta}, {Hou}, {Hou}, {L{\'e}pine}, {Li}, {Luo}, {Smith},
  {Wu}, {Yang}, {Yanny}, {Zhang}, \& {Zheng}}]{2015AJ....150....4C}
{Carlin}, J.~L., {Liu}, C., {Newberg}, H.~J., {et~al.} 2015, AJ, 150, 4

\bibitem[{{Casagrande} {et~al.}(2011){Casagrande}, {Sch{\"o}nrich}, {Asplund},
  {Cassisi}, {Ram{\'{\i}}rez}, {Mel{\'e}ndez}, {Bensby}, \&
  {Feltzing}}]{2011AA...530A.138C}
{Casagrande}, L., {Sch{\"o}nrich}, R., {Asplund}, M., {et~al.} 2011, \aap, 530,
  A138

\bibitem[{{Casey} {et~al.}(2016){Casey}, {Hawkins}, {Hogg}, {Ness},
  {Walter-Rix}, {Kordopatis}, {Kunder}, {Steinmetz}, {Koposov}, {Enke},
  {Sanders}, {Gilmore}, {Zwitter}, {Freeman}, {Casagrande}, {Matijevi{\v c}},
  {Seabroke}, {Bienaym{\'e}}, {Bland-Hawthorn}, {Gibson}, {Grebel}, {Helmi},
  {Munari}, {Navarro}, {Reid}, {Siebert}, \& {Wyse}}]{2016arXiv160902914C}
{Casey}, A.~R., {Hawkins}, K., {Hogg}, D.~W., {et~al.} 2016, ArXiv e-prints
  [\eprint[arXiv]{1609.02914}]

\bibitem[{{Chabrier}(2003)}]{2003PASP..115..763C}
{Chabrier}, G. 2003, \pasp, 115, 763

\bibitem[{{Cutri} \& {et al.}(2014)}]{2014yCat.2328....0C}
{Cutri}, R.~M. \& {et al.} 2014, VizieR Online Data Catalog, 2328

\bibitem[{{Demarque} {et~al.}(2004){Demarque}, {Woo}, {Kim}, \&
  {Yi}}]{2004ApJS..155..667D}
{Demarque}, P., {Woo}, J.-H., {Kim}, Y.-C., \& {Yi}, S.~K. 2004, \apjs, 155,
  667

\bibitem[{{Dotter} {et~al.}(2008){Dotter}, {Chaboyer}, {Jevremovi{\'c}},
  {Kostov}, {Baron}, \& {Ferguson}}]{2008ApJS..178...89D}
{Dotter}, A., {Chaboyer}, B., {Jevremovi{\'c}}, D., {et~al.} 2008, \apjs, 178,
  89

\bibitem[{{Gaia Collaboration} {et~al.}(2016){Gaia Collaboration}, {Brown},
  {Vallenari}, {Prusti}, {de Bruijne}, {Mignard}, {Drimmel}, \&
  {co-authors}}]{2016arXiv160904172G}
{Gaia Collaboration}, {Brown}, A.~G.~A., {Vallenari}, A., {et~al.} 2016, ArXiv
  e-prints [\eprint[arXiv]{1609.04172}]

\bibitem[{Gilmore(2015)}]{GAIA_ESO}
Gilmore, G. 2015, ESO Phase 3 Data Release Description

\bibitem[{{Gliese} \& {Jahrei{\ss}}(1991)}]{1991adc..rept.....G}
{Gliese}, W. \& {Jahrei{\ss}}, H. 1991, {Preliminary Version of the Third
  Catalogue of Nearby Stars}, Tech. rep.

\bibitem[{{Gontcharov}(2012)}]{2012AstL...38...87G}
{Gontcharov}, G.~A. 2012, Astronomy Letters, 38, 87

\bibitem[{{Green} {et~al.}(2015){Green}, {Schlafly}, {Finkbeiner}, {Rix},
  {Martin}, {Burgett}, {Draper}, {Flewelling}, {Hodapp}, {Kaiser}, {Kudritzki},
  {Magnier}, {Metcalfe}, {Price}, {Tonry}, \&
  {Wainscoat}}]{2015ApJ...810...25G}
{Green}, G.~M., {Schlafly}, E.~F., {Finkbeiner}, D.~P., {et~al.} 2015, \apj,
  810, 25

\bibitem[{{Hawkins} {et~al.}(2016){Hawkins}, {Masseron}, {Jofr{\'e}},
  {Gilmore}, {Elsworth}, \& {Hekker}}]{2016A&A...594A..43H}
{Hawkins}, K., {Masseron}, T., {Jofr{\'e}}, P., {et~al.} 2016, \aap, 594, A43

\bibitem[{{Ho} {et~al.}(2016){Ho}, {Ness}, {Hogg}, {Rix}, {Liu}, {Yang},
  {Zhang}, {Hou}, \& {Wang}}]{2016arXiv160200303H}
{Ho}, A.~Y.~Q., {Ness}, M.~K., {Hogg}, D.~W., {et~al.} 2016, ArXiv e-prints
  [\eprint[arXiv]{1602.00303}]

\bibitem[{{Howell} {et~al.}(2014){Howell}, {Sobeck}, {Haas}, {Still},
  {Barclay}, {Mullally}, {Troeltzsch}, {Aigrain}, {Bryson}, {Caldwell},
  {Chaplin}, {Cochran}, {Huber}, {Marcy}, {Miglio}, {Najita}, {Smith},
  {Twicken}, \& {Fortney}}]{2014PASP..126..398H}
{Howell}, S.~B., {Sobeck}, C., {Haas}, M., {et~al.} 2014, \pasp, 126, 398

\bibitem[{{Huber} {et~al.}(2016){Huber}, {Bryson}, {Haas}, {Barclay},
  {Barentsen}, {Howell}, {Sharma}, {Stello}, \&
  {Thompson}}]{2016ApJS..224....2H}
{Huber}, D., {Bryson}, S.~T., {Haas}, M.~R., {et~al.} 2016, \apjs, 224, 2

\bibitem[{{J{\o}rgensen} \& {Lindegren}(2005)}]{2005A&A...436..127J}
{J{\o}rgensen}, B.~R. \& {Lindegren}, L. 2005, \aap, 436, 127

\bibitem[{{Kjeldsen} \& {Bedding}(1995)}]{1995A&A...293...87K}
{Kjeldsen}, H. \& {Bedding}, T.~R. 1995, \aap, 293 [\eprint{astro-ph/9403015}]

\bibitem[{{Kordopatis} {et~al.}(2013){Kordopatis}, {Gilmore}, {Steinmetz},
  {Boeche}, {Seabroke}, {Siebert}, {Zwitter}, {Binney}, {de Laverny},
  {Recio-Blanco}, {Williams}, {Piffl}, {Enke}, {Roeser}, {Bijaoui}, {Wyse},
  {Freeman}, {Munari}, {Carrillo}, {Anguiano}, {Burton}, {Campbell}, {Cass},
  {Fiegert}, {Hartley}, {Parker}, {Reid}, {Ritter}, {Russell}, {Stupar},
  {Watson}, {Bienaym{\'e}}, {Bland-Hawthorn}, {Gerhard}, {Gibson}, {Grebel},
  {Helmi}, {Navarro}, {Conrad}, {Famaey}, {Faure}, {Just}, {Kos}, {Matijevi{\v
  c}}, {McMillan}, {Minchev}, {Scholz}, {Sharma}, {Siviero}, {de Boer}, \& {{\v
  Z}erjal}}]{2013AJ....146..134K}
{Kordopatis}, G., {Gilmore}, G., {Steinmetz}, M., {et~al.} 2013, \aj, 146, 134

\bibitem[{{Kroupa} \& {Weidner}(2003)}]{2003ApJ...598.1076K}
{Kroupa}, P. \& {Weidner}, C. 2003, \apj, 598, 1076

\bibitem[{{Kunder} {et~al.}(2016){Kunder}, {Kordopatis}, {Steinmetz},
  {Zwitter}, {McMillan}, {Casagrande}, {Enke}, {Wojno}, {Valentini},
  {Chiappini}, {Matijevic}, {Siviero}, {de Laverny}, {Recio-Blanco}, {Bijaoui},
  {Wyse}, {Binney}, {Grebel}, {Helmi}, {Jofre}, {Gilmore}, {Siebert}, {Famaey},
  {Bienayme}, {Gibson}, {Freeman}, {Navarro}, {Munari}, {Seabroke}, {Anguiano
  Jimenez}, {Reid}, {Bland-Hawthorn}, {Watson}, {Gerhard}, {Davies},
  {Elsworth}, {Lund}, {Miglio}, {Chaplin}, \& {Mosser}}]{2016arXiv160903210K}
{Kunder}, A., {Kordopatis}, G., {Steinmetz}, M., {et~al.} 2016, ArXiv e-prints
  [\eprint[arXiv]{1609.03210}]

\bibitem[{Leys {et~al.}(2013)Leys, Ley, Klein, Bernard, \& Licata}]{Leys2013}
Leys, C., Ley, C., Klein, O., Bernard, P., \& Licata, L. 2013

\bibitem[{{Lindegren} {et~al.}(2016){Lindegren}, {Lammers}, {Bastian},
  {Hern{\'a}ndez}, {Klioner}, {Hobbs}, {Bombrun}, {Michalik}, {Ramos-Lerate},
  {Butkevich}, {Comoretto}, {Joliet}, {Holl}, {Hutton}, {Parsons},
  {Steidelm{\"u}ller}, {Abbas}, {Altmann}, {Andrei}, {Anton}, {Bach},
  {Barache}, {Becciani}, {Berthier}, {Bianchi}, {Biermann}, {Bouquillon},
  {Bourda}, {Br{\"u}semeister}, {Bucciarelli}, {Busonero}, {Carlucci},
  {Casta{\~n}eda}, {Charlot}, {Clotet}, {Crosta}, {Davidson}, {de Felice},
  {Drimmel}, {Fabricius}, {Fienga}, {Figueras}, {Fraile}, {Gai}, {Garralda},
  {Geyer}, {Gonz{\'a}lez-Vidal}, {Guerra}, {Hambly}, {Hauser}, {Jordan},
  {Lattanzi}, {Lenhardt}, {Liao}, {L{\"o}ffler}, {McMillan}, {Mignard}, {Mora},
  {Morbidelli}, {Portell}, {Riva}, {Sarasso}, {Serraller}, {Siddiqui}, {Smart},
  {Spagna}, {Stampa}, {Steele}, {Taris}, {Torra}, {van Reeven}, {Vecchiato},
  {Zschocke}, {de Bruijne}, {Gracia}, {Raison}, {Lister}, {Marchant},
  {Messineo}, {Soffel}, {Osorio}, {de Torres}, \&
  {O'Mullane}}]{2016arXiv160904303L}
{Lindegren}, L., {Lammers}, U., {Bastian}, U., {et~al.} 2016, ArXiv e-prints
  [\eprint[arXiv]{1609.04303}]

\bibitem[{{Luo} {et~al.}(2015){Luo}, {Zhao}, {Zhao}, {Deng}, {Liu}, {Jing},
  {Wang}, {Zhang}, {Shi}, {Cui}, {Chu}, {Li}, {Bai}, {Wu}, {Cai}, {Cao}, {Cao},
  {Carlin}, {Chen}, {Chen}, {Chen}, {Chen}, {Chen}, {Chen}, {Chen},
  {Christlieb}, {Chu}, {Cui}, {Dong}, {Du}, {Fan}, {Feng}, {Fu}, {Gao}, {Gong},
  {Gu}, {Guo}, {Han}, {He}, {Hou}, {Hou}, {Hou}, {Hu}, {Hu}, {Hu}, {Huo},
  {Jia}, {Jiang}, {Jiang}, {Jiang}, {Jin}, {Kong}, {Kong}, {Lei}, {Li}, {Li},
  {Li}, {Li}, {Li}, {Li}, {Li}, {Li}, {Li}, {Li}, {Li}, {Li}, {Liang}, {Lin},
  {Liu}, {Liu}, {Liu}, {Liu}, {Lu}, {Luo}, {Mao}, {Newberg}, {Ni}, {Qi}, {Qi},
  {Shen}, {Shi}, {Song}, {Song}, {Su}, {Su}, {Tang}, {Tao}, {Tian}, {Wang},
  {Wang}, {Wang}, {Wang}, {Wang}, {Wang}, {Wang}, {Wang}, {Wang}, {Wang},
  {Wang}, {Wang}, {Wang}, {Wang}, {Wang}, {Wang}, {Wang}, {Wang}, {Wang},
  {Wang}, {Wei}, {Wei}, {Wu}, {Wu}, {Wu}, {Wu}, {Xing}, {Xu}, {Xu}, {Xu},
  {Yan}, {Yang}, {Yang}, {Yang}, {Yang}, {Yao}, {Yu}, {Yuan}, {Yuan}, {Yuan},
  {Yuan}, {Zhai}, {Zhang}, {Zhang}, {Zhang}, {Zhang}, {Zhang}, {Zhang},
  {Zhang}, {Zhang}, {Zhao}, {Zhou}, {Zhou}, {Zhu}, {Zhu}, {Zou}, \&
  {Zuo}}]{2015RAA....15.1095L}
{Luo}, A.-L., {Zhao}, Y.-H., {Zhao}, G., {et~al.} 2015, Research in Astronomy
  and Astrophysics, 15, 1095

\bibitem[{{Majewski} {et~al.}(2000){Majewski}, {Ostheimer}, {Kunkel}, \&
  {Patterson}}]{2000AJ....120.2550M}
{Majewski}, S.~R., {Ostheimer}, J.~C., {Kunkel}, W.~E., \& {Patterson}, R.~J.
  2000, \aj, 120, 2550

\bibitem[{{Majewski} {et~al.}(2011){Majewski}, {Zasowski}, \&
  {Nidever}}]{2011ApJ...739...25M}
{Majewski}, S.~R., {Zasowski}, G., \& {Nidever}, D.~L. 2011, \apj, 739, 25

\bibitem[{{Martell} {et~al.}(2016){Martell}, {Sharma}, {Buder}, {Duong},
  {Schlesinger}, {Simpson}, {Lind}, {Ness}, {Marshall}, {Asplund},
  {Bland-Hawthorn}, {Casey}, {De Silva}, {Freeman}, {Kos}, {Lin}, {Zucker},
  {Zwitter}, {Anguiano}, {Bacigalupo}, {Carollo}, {Casagrande}, {Da Costa},
  {Horner}, {Huber}, {Hyde}, {Kafle}, {Lewis}, {Nataf}, {Stello}, {Tinney},
  {Watson}, \& {Wittenmyer}}]{GALAH}
{Martell}, S., {Sharma}, S., {Buder}, S., {et~al.} 2016, ArXiv e-prints
  [\eprint[arXiv]{1609.02822}]

\bibitem[{{Martig} {et~al.}(2016){Martig}, {Fouesneau}, {Rix}, {Ness},
  {M{\'e}sz{\'a}ros}, {Garc{\'{\i}}a-Hern{\'a}ndez}, {Pinsonneault},
  {Serenelli}, {Silva Aguirre}, \& {Zamora}}]{2016MNRAS.456.3655M}
{Martig}, M., {Fouesneau}, M., {Rix}, H.-W., {et~al.} 2016, \mnras, 456, 3655

\bibitem[{{Ness} {et~al.}(2015){Ness}, {Hogg}, {Rix}, {Ho}, \&
  {Zasowski}}]{2015ApJ...808...16N}
{Ness}, M., {Hogg}, D.~W., {Rix}, H.-W., {Ho}, A.~Y.~Q., \& {Zasowski}, G.
  2015, \apj, 808, 16

\bibitem[{{Perryman} {et~al.}(1997){Perryman}, {Lindegren}, {Kovalevsky},
  {Hoeg}, {Bastian}, {Bernacca}, {Cr{\'e}z{\'e}}, {Donati}, {Grenon},
  {Grewing}, {van Leeuwen}, {van der Marel}, {Mignard}, {Murray}, {Le Poole},
  {Schrijver}, {Turon}, {Arenou}, {Froeschl{\'e}}, \&
  {Petersen}}]{1997A&A...323L..49P}
{Perryman}, M.~A.~C., {Lindegren}, L., {Kovalevsky}, J., {et~al.} 1997, \aap,
  323, L49

\bibitem[{{Pietrinferni} {et~al.}(2009){Pietrinferni}, {Cassisi}, {Salaris},
  {Percival}, \& {Ferguson}}]{2009ApJ...697..275P}
{Pietrinferni}, A., {Cassisi}, S., {Salaris}, M., {Percival}, S., \&
  {Ferguson}, J.~W. 2009, \apj, 697, 275

\bibitem[{{Pinsonneault} {et~al.}(2014){Pinsonneault}, {Elsworth}, {Epstein},
  {Hekker}, {M{\'e}sz{\'a}ros}, {Chaplin}, {Johnson}, {Garc{\'{\i}}a},
  {Holtzman}, {Mathur}, {Garc{\'{\i}}a P{\'e}rez}, {Silva Aguirre}, {Girardi},
  {Basu}, {Shetrone}, {Stello}, {Allende Prieto}, {An}, {Beck}, {Beers},
  {Bizyaev}, {Bloemen}, {Bovy}, {Cunha}, {De Ridder}, {Frinchaboy},
  {Garc{\'{\i}}a-Hern{\'a}ndez}, {Gilliland}, {Harding}, {Hearty}, {Huber},
  {Ivans}, {Kallinger}, {Majewski}, {Metcalfe}, {Miglio}, {Mosser}, {Muna},
  {Nidever}, {Schneider}, {Serenelli}, {Smith}, {Tayar}, {Zamora}, \&
  {Zasowski}}]{2014ApJS..215...19P}
{Pinsonneault}, M.~H., {Elsworth}, Y., {Epstein}, C., {et~al.} 2014, \apjs,
  215, 19

\bibitem[{{Rauer} {et~al.}(2014){Rauer}, {Catala}, {Aerts}, {Appourchaux},
  {Benz}, {Brandeker}, {Christensen-Dalsgaard}, {Deleuil}, {Gizon}, {Goupil},
  {G{\"u}del}, {Janot-Pacheco}, {Mas-Hesse}, {Pagano}, {Piotto}, {Pollacco},
  {Santos}, {Smith}, {Su{\'a}rez}, {Szab{\'o}}, {Udry}, {Adibekyan}, {Alibert},
  {Almenara}, {Amaro-Seoane}, {Eiff}, {Asplund}, {Antonello}, {Barnes},
  {Baudin}, {Belkacem}, {Bergemann}, {Bihain}, {Birch}, {Bonfils}, {Boisse},
  {Bonomo}, {Borsa}, {Brand{\~a}o}, {Brocato}, {Brun}, {Burleigh}, {Burston},
  {Cabrera}, {Cassisi}, {Chaplin}, {Charpinet}, {Chiappini}, {Church},
  {Csizmadia}, {Cunha}, {Damasso}, {Davies}, {Deeg}, {D{\'{\i}}az}, {Dreizler},
  {Dreyer}, {Eggenberger}, {Ehrenreich}, {Eigm{\"u}ller}, {Erikson}, {Farmer},
  {Feltzing}, {de Oliveira Fialho}, {Figueira}, {Forveille}, {Fridlund},
  {Garc{\'{\i}}a}, {Giommi}, {Giuffrida}, {Godolt}, {Gomes da Silva},
  {Granzer}, {Grenfell}, {Grotsch-Noels}, {G{\"u}nther}, {Haswell}, {Hatzes},
  {H{\'e}brard}, {Hekker}, {Helled}, {Heng}, {Jenkins}, {Johansen},
  {Khodachenko}, {Kislyakova}, {Kley}, {Kolb}, {Krivova}, {Kupka}, {Lammer},
  {Lanza}, {Lebreton}, {Magrin}, {Marcos-Arenal}, {Marrese}, {Marques},
  {Martins}, {Mathis}, {Mathur}, {Messina}, {Miglio}, {Montalban}, {Montalto},
  {Monteiro}, {Moradi}, {Moravveji}, {Mordasini}, {Morel}, {Mortier},
  {Nascimbeni}, {Nelson}, {Nielsen}, {Noack}, {Norton}, {Ofir}, {Oshagh},
  {Ouazzani}, {P{\'a}pics}, {Parro}, {Petit}, {Plez}, {Poretti}, {Quirrenbach},
  {Ragazzoni}, {Raimondo}, {Rainer}, {Reese}, {Redmer}, {Reffert},
  {Rojas-Ayala}, {Roxburgh}, {Salmon}, {Santerne}, {Schneider}, {Schou},
  {Schuh}, {Schunker}, {Silva-Valio}, {Silvotti}, {Skillen}, {Snellen}, {Sohl},
  {Sousa}, {Sozzetti}, {Stello}, {Strassmeier}, {{\v S}vanda}, {Szab{\'o}},
  {Tkachenko}, {Valencia}, {Van Grootel}, {Vauclair}, {Ventura}, {Wagner},
  {Walton}, {Weingrill}, {Werner}, {Wheatley}, \&
  {Zwintz}}]{2014ExA....38..249R}
{Rauer}, H., {Catala}, C., {Aerts}, C., {et~al.} 2014, Experimental Astronomy,
  38, 249

\bibitem[{{Ricker} {et~al.}(2014){Ricker}, {Winn}, {Vanderspek}, {Latham},
  {Bakos}, {Bean}, {Berta-Thompson}, {Brown}, {Buchhave}, {Butler}, {Butler},
  {Chaplin}, {Charbonneau}, {Christensen-Dalsgaard}, {Clampin}, {Deming},
  {Doty}, {De Lee}, {Dressing}, {Dunham}, {Endl}, {Fressin}, {Ge}, {Henning},
  {Holman}, {Howard}, {Ida}, {Jenkins}, {Jernigan}, {Johnson}, {Kaltenegger},
  {Kawai}, {Kjeldsen}, {Laughlin}, {Levine}, {Lin}, {Lissauer}, {MacQueen},
  {Marcy}, {McCullough}, {Morton}, {Narita}, {Paegert}, {Palle}, {Pepe},
  {Pepper}, {Quirrenbach}, {Rinehart}, {Sasselov}, {Sato}, {Seager},
  {Sozzetti}, {Stassun}, {Sullivan}, {Szentgyorgyi}, {Torres}, {Udry}, \&
  {Villasenor}}]{2014SPIE.9143E..20R}
{Ricker}, G.~R., {Winn}, J.~N., {Vanderspek}, R., {et~al.} 2014, in \procspie,
  Vol. 9143, Space Telescopes and Instrumentation 2014: Optical, Infrared, and
  Millimeter Wave, 914320

\bibitem[{{Rodrigues} {et~al.}(2014){Rodrigues}, {Girardi}, {Miglio},
  {Bossini}, {Bovy}, {Epstein}, {Pinsonneault}, {Stello}, {Zasowski}, {Prieto},
  {Chaplin}, {Hekker}, {Johnson}, {M{\'e}sz{\'a}ros}, {Mosser}, {Anders},
  {Basu}, {Beers}, {Chiappini}, {da Costa}, {Elsworth}, {Garc{\'{\i}}a},
  {P{\'e}rez}, {Hearty}, {Maia}, {Majewski}, {Mathur}, {Montalb{\'a}n},
  {Nidever}, {Santiago}, {Schultheis}, {Serenelli}, \&
  {Shetrone}}]{2014MNRAS.445.2758R}
{Rodrigues}, T.~S., {Girardi}, L., {Miglio}, A., {et~al.} 2014, \mnras, 445,
  2758

\bibitem[{{Roeser} {et~al.}(2010){Roeser}, {Demleitner}, \&
  {Schilbach}}]{2010AJ....139.2440R}
{Roeser}, S., {Demleitner}, M., \& {Schilbach}, E. 2010, \aj, 139, 2440

\bibitem[{{Salaris} {et~al.}(1993){Salaris}, {Chieffi}, \&
  {Straniero}}]{1993ApJ...414..580S}
{Salaris}, M., {Chieffi}, A., \& {Straniero}, O. 1993, \apj, 414, 580

\bibitem[{{Sale} {et~al.}(2014){Sale}, {Drew}, {Barentsen}, {Farnhill},
  {Raddi}, {Barlow}, {Eisl{\"o}ffel}, {Vink}, {Rodr{\'{\i}}guez-Gil}, \&
  {Wright}}]{2014MNRAS.443.2907S}
{Sale}, S.~E., {Drew}, J.~E., {Barentsen}, G., {et~al.} 2014, \mnras, 443, 2907

\bibitem[{{S{\'a}nchez-Bl{\'a}zquez} {et~al.}(2006){S{\'a}nchez-Bl{\'a}zquez},
  {Peletier}, {Jim{\'e}nez-Vicente}, {Cardiel}, {Cenarro},
  {Falc{\'o}n-Barroso}, {Gorgas}, {Selam}, \& {Vazdekis}}]{2006MNRAS.371..703S}
{S{\'a}nchez-Bl{\'a}zquez}, P., {Peletier}, R.~F., {Jim{\'e}nez-Vicente}, J.,
  {et~al.} 2006, \mnras, 371, 703

\bibitem[{{Schlegel} {et~al.}(1998){Schlegel}, {Finkbeiner}, \&
  {Davis}}]{1998ApJ...500..525S}
{Schlegel}, D.~J., {Finkbeiner}, D.~P., \& {Davis}, M. 1998, \apj, 500, 525

\bibitem[{{Sch{\"o}nrich} {et~al.}(2012){Sch{\"o}nrich}, {Binney}, \&
  {Asplund}}]{2012MNRAS.420.1281S}
{Sch{\"o}nrich}, R., {Binney}, J., \& {Asplund}, M. 2012, \mnras, 420, 1281

\bibitem[{{SDSS Collaboration} {et~al.}(2016){SDSS Collaboration}, {Albareti},
  {Allende Prieto}, {Almeida}, {Anders}, {Anderson}, {Andrews},
  {Aragon-Salamanca}, {Argudo-Fernandez}, {Armengaud}, \&
  et~al.}]{2016arXiv160802013S}
{SDSS Collaboration}, {Albareti}, F.~D., {Allende Prieto}, C., {et~al.} 2016,
  ArXiv e-prints [\eprint[arXiv]{1608.02013}]

\bibitem[{{Silva Aguirre} \& {Serenelli}(2016)}]{2016AN....337..823S}
{Silva Aguirre}, V. \& {Serenelli}, A.~M. 2016, Astronomische Nachrichten, 337,
  823

\bibitem[{{Skrutskie} {et~al.}(2006){Skrutskie}, {Cutri}, {Stiening},
  {Weinberg}, {Schneider}, {Carpenter}, {Beichman}, {Capps}, {Chester},
  {Elias}, {Huchra}, {Liebert}, {Lonsdale}, {Monet}, {Price}, {Seitzer},
  {Jarrett}, {Kirkpatrick}, {Gizis}, {Howard}, {Evans}, {Fowler}, {Fullmer},
  {Hurt}, {Light}, {Kopan}, {Marsh}, {McCallon}, {Tam}, {Van Dyk}, \&
  {Wheelock}}]{2006AJ....131.1163S}
{Skrutskie}, M.~F., {Cutri}, R.~M., {Stiening}, R., {et~al.} 2006, \aj, 131,
  1163

\bibitem[{{Soderblom}(2010)}]{2010ARA&A..48..581S}
{Soderblom}, D.~R. 2010, \araa, 48, 581

\bibitem[{{Stassun} \& {Torres}(2016)}]{2016arXiv160905390S}
{Stassun}, K.~G. \& {Torres}, G. 2016, ArXiv e-prints
  [\eprint[arXiv]{1609.05390}]

\bibitem[{{Valenti} \& {Piskunov}(2012)}]{2012ascl.soft02013V}
{Valenti}, J.~A. \& {Piskunov}, N. 2012, {SME: Spectroscopy Made Easy},
  Astrophysics Source Code Library

\bibitem[{{Wang} {et~al.}(2016{\natexlab{a}}){Wang}, {Shi}, {Zhao}, {Zhang},
  {Huo}, {Zhang}, {Chen}, {Wu}, {Zhang}, \& {Hou}}]{2016MNRAS.456..672W}
{Wang}, J., {Shi}, J., {Zhao}, Y., {et~al.} 2016{\natexlab{a}}, \mnras, 456,
  672

\bibitem[{{Wang} {et~al.}(2016{\natexlab{b}}){Wang}, {Wang}, {Wu}, {Zhao},
  {Li}, {Luo}, {Liu}, {Zhang}, {Hou}, {Wang}, \& {Cao}}]{2016AJ....152....6W}
{Wang}, L., {Wang}, W., {Wu}, Y., {et~al.} 2016{\natexlab{b}}, \aj, 152, 6

\bibitem[{{Wenger} {et~al.}(2000){Wenger}, {Ochsenbein}, {Egret}, {Dubois},
  {Bonnarel}, {Borde}, {Genova}, {Jasniewicz}, {Lalo{\"e}}, {Lesteven}, \&
  {Monier}}]{2000A&AS..143....9W}
{Wenger}, M., {Ochsenbein}, F., {Egret}, D., {et~al.} 2000, \aaps, 143, 9

\bibitem[{{Woolley} {et~al.}(1970){Woolley}, {Epps}, {Penston}, \&
  {Pocock}}]{1970ROAn....5.....W}
{Woolley}, R., {Epps}, E.~A., {Penston}, M.~J., \& {Pocock}, S.~B. 1970, Royal
  Observatory Annals, 5

\bibitem[{{Worley} {et~al.}(2016){Worley}, {de Laverny}, {Recio-Blanco},
  {Hill}, \& {Bijaoui}}]{AMBRE}
{Worley}, C.~C., {de Laverny}, P., {Recio-Blanco}, A., {Hill}, V., \&
  {Bijaoui}, A. 2016, \aap, 591, A81

\bibitem[{{Xiang} {et~al.}(2017){Xiang}, {Liu}, {Yuan}, {Huo}, {Huang}, {Wang},
  {Chen}, {Ren}, {Zhang}, {Tian}, {Yang}, {Shi}, {Zhao}, {Li}, {Zhao}, {Cui},
  {Li}, {Hou}, {Zhang}, {Zhang}, {Wang}, {Wu}, {Cao}, {Yan}, {Yan}, {Luo},
  {Zhang}, {Bai}, {Yuan}, {Dong}, {Lei}, \& {Li}}]{2017arXiv170105409X}
{Xiang}, M., {Liu}, X., {Yuan}, H., {et~al.} 2017, ArXiv e-prints
  [\eprint[arXiv]{1701.05409}]

\bibitem[{{Yanny} {et~al.}(2009){Yanny}, {Rockosi}, {Newberg}, {Knapp},
  {Adelman-McCarthy}, {Alcorn}, {Allam}, {Allende Prieto}, {An}, {Anderson},
  {Anderson}, {Bailer-Jones}, {Bastian}, {Beers}, {Bell}, {Belokurov},
  {Bizyaev}, {Blythe}, {Bochanski}, {Boroski}, {Brinchmann}, {Brinkmann},
  {Brewington}, {Carey}, {Cudworth}, {Evans}, {Evans}, {Gates}, {G{\"a}nsicke},
  {Gillespie}, {Gilmore}, {Nebot Gomez-Moran}, {Grebel}, {Greenwell}, {Gunn},
  {Jordan}, {Jordan}, {Harding}, {Harris}, {Hendry}, {Holder}, {Ivans},
  {Ivezi{\v c}}, {Jester}, {Johnson}, {Kent}, {Kleinman}, {Kniazev},
  {Krzesinski}, {Kron}, {Kuropatkin}, {Lebedeva}, {Lee}, {French Leger},
  {L{\'e}pine}, {Levine}, {Lin}, {Long}, {Loomis}, {Lupton}, {Malanushenko},
  {Malanushenko}, {Margon}, {Martinez-Delgado}, {McGehee}, {Monet}, {Morrison},
  {Munn}, {Neilsen}, {Nitta}, {Norris}, {Oravetz}, {Owen}, {Padmanabhan},
  {Pan}, {Peterson}, {Pier}, {Platson}, {Re Fiorentin}, {Richards}, {Rix},
  {Schlegel}, {Schneider}, {Schreiber}, {Schwope}, {Sibley}, {Simmons},
  {Snedden}, {Allyn Smith}, {Stark}, {Stauffer}, {Steinmetz}, {Stoughton},
  {SubbaRao}, {Szalay}, {Szkody}, {Thakar}, {Sivarani}, {Tucker}, {Uomoto},
  {Vanden Berk}, {Vidrih}, {Wadadekar}, {Watters}, {Wilhelm}, {Wyse}, {Yarger},
  \& {Zucker}}]{2009AJ....137.4377Y}
{Yanny}, B., {Rockosi}, C., {Newberg}, H.~J., {et~al.} 2009, \aj, 137, 4377

\bibitem[{{Yuan} {et~al.}(2015){Yuan}, {Liu}, {Huo}, {Xiang}, {Huang}, {Chen},
  {Zhang}, {Sun}, {Wang}, {Zhang}, {Zhao}, {Luo}, {Shi}, {Li}, {Yuan}, {Dong},
  {Li}, {Hou}, \& {Zhang}}]{2015MNRAS.448..855Y}
{Yuan}, H.~B., {Liu}, X.~W., {Huo}, Z.~Y., {et~al.} 2015, \mnras, 448, 855

\bibitem[{{Yuan} {et~al.}(2013){Yuan}, {Liu}, \& {Xiang}}]{Extinction}
{Yuan}, H.~B., {Liu}, X.~W., \& {Xiang}, M.~S. 2013, \mnras, 430, 2188

\bibitem[{{Yuan} {et~al.}(2014){Yuan}, {Liu}, {Xiang}, {Huo}, {Zhang}, {Huang},
  \& {Zhang}}]{2014IAUS..298..240Y}
{Yuan}, H.-B., {Liu}, X.-W., {Xiang}, M.-S., {et~al.} 2014, in IAU Symposium,
  Vol. 298, Setting the scene for Gaia and LAMOST, ed. S.~{Feltzing},
  G.~{Zhao}, N.~A. {Walton}, \& P.~{Whitelock}, 240--245

\bibitem[{{Zacharias} {et~al.}(2013){Zacharias}, {Finch}, {Girard}, {Henden},
  {Bartlett}, {Monet}, \& {Zacharias}}]{2013AJ....145...44Z}
{Zacharias}, N., {Finch}, C.~T., {Girard}, T.~M., {et~al.} 2013, \aj, 145, 44

\bibitem[{{Zhang} {et~al.}(2014){Zhang}, {Liu}, {Yuan}, {Zhao}, {Yao}, {Zhang},
  {Xiang}, \& {Huang}}]{2014RAA....14..456Z}
{Zhang}, H.-H., {Liu}, X.-W., {Yuan}, H.-B., {et~al.} 2014, Research in
  Astronomy and Astrophysics, 14, 456

\bibitem[{{Zwitter} {et~al.}(2010){Zwitter}, {Matijevi{\v c}}, {Breddels},
  {Smith}, {Helmi}, {Munari}, {Bienaym{\'e}}, {Binney}, {Bland-Hawthorn},
  {Boeche}, {Brown}, {Campbell}, {Freeman}, {Fulbright}, {Gibson}, {Gilmore},
  {Grebel}, {Navarro}, {Parker}, {Seabroke}, {Siebert}, {Siviero}, {Steinmetz},
  {Watson}, {Williams}, \& {Wyse}}]{2010A&A...522A..54Z}
{Zwitter}, T., {Matijevi{\v c}}, G., {Breddels}, M.~A., {et~al.} 2010, \aap,
  522, A54

\end{thebibliography}

\appendix
\section{Splitting the PDF into unimodal sub-PDFs}\label{sec:split_uspdf}
Here we describe a method used to split complex PDFs into a set of unimodal sub-PDFs.

First, we produced a histogram for logarithm of stellar masses of models of the same evolutionary stage (weighted by $W_j$, see \autoref{eq:model_weight}). Then, we detected local minima and maxima of this histogram. Local minima (or maxima) are defined as locations of bins that have lower (higher) value $h_i$ than all other bins within the window: i.e. $h_i = min\{h_j, i-n \leq j \leq i+n\}$ for a local minimum and $h_i = max\{h_j, i-n \leq j \leq i+n\}$ for a local maximum. Window size $n$ was taken to be 3 for maxima and 2 for minima. Differences in window sizes are caused by the need to locate minima with high precision and to avoid too many maxima in noisy data. Formally, it is possible to have more than one local minimum between two local maxima; we split only by the lowest of them in this case. We split the sample at positions of local minima that are lower than 0.75 times the value of the smallest of the two enclosing maxima. We thus could have one or several USPDFs, for each evolutionary stage. 

We chose the histogram in logarithm of mass to split the multimodal PDFs as the logarithmic scale is close to linear one around $1 \Msun$, but gives a smoother histogram for high-mass stars. Mass is a better choice to split the PDF because values of log(age)s are quantised by construction of the isochrones, and distances are much less sensitive to evolutionary stage. 

\section{Distributions used in the paper}\label{app:functions}
Here we give definitions for some functions used in the paper. 

We define $\phi(x)=\frac{1}{\sqrt{2 \pi}}\exp\left(-\frac{1}{2}x^2\right)$ as the PDF of the standard normal distribution and $\Phi(x)$ as it's cumulative distribution. This PDF is designated as a Gaussian in the text.

Introducing $\xi=\frac{x-\alpha}{\sigma}$ and $Z=\Phi(\frac{b-\alpha}{\sigma})-\Phi(\frac{a-\alpha}{\sigma})$, we have for a truncated Gaussian
\begin{equation}
f(x,\alpha,\sigma, a,b) = \frac{\phi(\xi)}{\sigma Z},
\end{equation}
if $a < x < b$ and $f(x;\alpha,\sigma, a,b) = 0$ otherwise. Here $\alpha$ is a location, $\sigma$ - scale, $a, b$ - lower and upper limits.

For skewed Gaussian with a shape parameter $s$
\begin{equation}
f(x,\alpha,\sigma, s) = \frac{2}{\sigma} \phi(\xi) \Phi(s \xi).
\end{equation}

We use the definition of the truncated \stud
\begin{equation}
 f_t(x, \nu) = \frac{\Gamma \left(\frac{\nu+1}{2} \right)} {\sqrt{\nu\pi}\,\Gamma \left(\frac{\nu}{2} \right)} \left(1+\frac{x^2}{\nu} \right)^{-\frac{\nu+1}{2}},
\end{equation} 
where $\nu$ is the ``number of degrees of freedom'' (which can be an arbitrary real number here). Again, $\Phi_t(x, \nu)$ is the cumulative distribution function. 

A modified truncated exponential distribution with lower and upper limits $a$ and $b$, respectively is defined as
\begin{equation}
 f_{\textrm{exp}}(x, \alpha, \sigma, a, b) = 
 \begin{cases}
   C e^{-\frac{|x-\alpha|}{\sigma}}, & \textrm{if}\ a < x < b \\
   0,& \textrm{otherwise.}
 \end{cases}
\end{equation} 
Here, $C$ is the normalisation constant, so that $\int_a^b f_{\textrm{exp}}(x, \alpha, \sigma, a, b) = 1$.

For a truncated Student's t-distribution with lower and upper limits $a$ and $b$, respectively, we define $\xi=\frac{x-\alpha}{\sigma}$ and $Z_t=\Phi_t(\frac{b-\alpha}{\sigma}, \nu)-\Phi_t(\frac{a-\alpha}{\sigma}, \nu)$. Then for the PDF we have
\begin{equation}
 f_{\textrm{truncated-t}}(x, \alpha, \sigma, a, b) = \frac{f_t(\xi, \nu)}{Z_t}.
\end{equation} 

\end{document}